\documentclass[11pt,aps,prd,preprint,superscriptaddress,showpacs,nofootinbib]{revtex4}
\pdfoutput=1
\usepackage{revsymb}
\usepackage{amssymb}
\usepackage{amsmath}
\usepackage{amsfonts}

\usepackage{placeins}
\usepackage[pdftex]{graphicx}
\usepackage{float}
\usepackage{subfigure}
\usepackage{booktabs}
\usepackage{multirow}
\DeclareGraphicsExtensions{.pdf,.png,.jpg} % busca en este orden!

\newcommand{\be}{\begin{equation}}
\newcommand{\ee}{\end{equation}}
\newcommand{\ben}{\begin{eqnarray}}
\newcommand{\een}{\end{eqnarray}}

\newcommand{\n}{\label}

\begin{document}
\title{Does a generalized Chaplygin gas correctly describe\\ the cosmological dark
sector?}
\author{R.F. vom Marttens\footnote{E-mail: rodrigovonmarttens@gmail.com}}
\affiliation{Universidade Federal do Esp\'{\i}rito Santo,
Departamento
de F\'{\i}sica\\
Av. Fernando Ferrari, 514, Campus de Goiabeiras, CEP 29075-910,
Vit\'oria, Esp\'{\i}rito Santo, Brazil}
\author{L. Casarini \footnote{E-mail: casarini.astro@gmail.com}}
\affiliation{Universidade Federal do Esp\'{\i}rito Santo,
Departamento
de F\'{\i}sica\\
Av. Fernando Ferrari, 514, Campus de Goiabeiras, CEP 29075-910,
Vit\'oria, Esp\'{\i}rito Santo, Brazil}
\affiliation{Institute of Theoretical Astrophysics, University of Oslo, 0315 Oslo, Norway}
\author{\\ W. Zimdahl\footnote{E-mail: winfried.zimdahl@pq.cnpq.br}}
\affiliation{Universidade Federal do Esp\'{\i}rito Santo,
Departamento
de F\'{\i}sica\\
Av. Fernando Ferrari, 514, Campus de Goiabeiras, CEP 29075-910,
Vit\'oria, Esp\'{\i}rito Santo, Brazil}
\author{W.S. Hip\'{o}lito-Ricaldi\footnote{E-mail: wiliam.ricaldi@ufes.br}}
\affiliation{Universidade Federal do Esp\'{\i}rito Santo, Departamento de Ci\^encias  Naturais\\
Rodovia BR 101 Norte, km. 60, CEP 29932-540,
S\~ao Mateus, Esp\'{\i}rito Santo, Brazil}
\author{D.F. Mota\footnote{E-mail: d.f.mota@astro.uio.no}}
\affiliation{Institute of Theoretical Astrophysics, University of Oslo, 0315 Oslo, Norway}

\date{\today}

\begin{abstract}
Yes, but only for a parameter value that makes it almost
coincide with the standard model.
We reconsider the cosmological dynamics of a generalized Chaplygin gas (gCg) which is  split into
a cold dark matter (CDM) part and a dark energy (DE) component with constant equation of state.
This model, which implies a specific interaction between CDM and DE, has a $\Lambda$CDM limit and provides the basis
for studying deviations from the latter. Including  matter and radiation,  we use the (modified) CLASS code \cite{class} to construct the CMB and matter power spectra in order to search for a gCg-based concordance
model that is in agreement with the SNIa data from the JLA sample and with recent Planck data. The results   reveal that the gCg parameter $\alpha$ is restricted
to $|\alpha|\lesssim 0.05$, i.e., to values very close to the $\Lambda$CDM limit $\alpha =0$. This excludes, in particular,
models in which DE decays linearly with the Hubble rate.
\end{abstract}

%\pacs{98.80.-k, 95.35.+d, 95.36.+x}

\maketitle
% * <winfried.zimdahl@gmail.com> 2016-10-14T19:42:35.850Z:
%
% ^.
% * <winfried.zimdahl@gmail.com> 2016-11-04T20:44:21.475Z:
%
% ^.
\section{Introduction}

A large part of the current cosmological literature is devoted both to a theoretical understanding of the $\Lambda$CDM model and to its observational verification. While it has become the status of a standard model it relies on the assumption of a dark sector which is far from being understood physically. Given its simplicity, it is considered very successful observationally, no competing model is doing better at the moment, but there remain also tensions \cite{buchert15}.
It is of ongoing interest therefore to check the status of the $\Lambda$CDM model by modifying its basic assumption and to test the observational consequences of such modifications.

One line of research that has been followed in this context relies on the dynamics of Chaplygin gases
\cite{Chaplygin}.  The Chaplygin gas in its original form, characterized by an equation of state (EoS)
$p = - \frac{A}{\rho}$, was applied to cosmology in  \cite{moschella} followed by \cite{julio,bilic}. Here, $A$ is a strictly positive constant, $p$ is the pressure and $\rho$ is the energy density.
Its relation to higher-dimensional theories was pointed out in \cite{Jackiw}.
A phenomenological generalization to an  EoS
\be
p=-\dfrac{A}{\rho^{\alpha}},
\n{chaplygin}
\ee
with a constant $\alpha>-1$  was introduced in \cite{berto}, where also its relation to a scalar-field Lagrangian
 of a generalized Born-Infeld type was clarified. 
For $\alpha =1$ the original Chaplygin gas is recovered, the case $\alpha =0$ is related to the $\Lambda$CDM model.
A Chaplygin gas has the appealing feature  that it allows for a unified description of the dark sector.
Its  energy density  changes smoothly from that of nonrelativistic matter at early times to an almost constant value in the far future.
Thus it may  interpolate between an early phase of decelerated
expansion, necessary for structure formation to occur, and a later period in which
this substratum acts similarly as a cosmological constant, giving
rise to an accelerated expansion.
While the mentioned unifying aspects seem to offer a conceptual advantage compared with other approaches, one faces the problem that
a successful description of structure
formation requires a separation of the observable pressureless matter component.
At first sight this seems to be a step back. However, since for the gCg the total energy density is analytically known
it is possible to identify the coupling of this separated matter part to the remaining part that plays the role of DE
and completes the overall Chaplygin gas.
Cosmological models based on the dynamics of generalized Chaplygin gases have attracted considerable interest \cite{oliver08,oliver09,gcgjulio1,ioav,mota2,rrrr,VDF1,VDF2,gcgjulio2,nadgcg,wands,saulo14}.
But this type of models had temporarily been seen as disfavored since in their adiabatic version they predict unobserved oscillations and/or instabilities in the matter power spectrum \cite{ioav}. It turned out, however, that nonadiabatic perturbations may remove such unwanted features \cite{rrrr,VDF1,VDF2}.
Generalized Chaplygin gas models share similarities with decaying vacuum models
(see, e.g.,\cite{oesertaha,oesertaha87,mota1,berto2,Freese,WangMeng,WangGongAbd,Borges,pertsaulo})
which result as special  cases if the constant EoS parameter of the DE component is chosen to be $-1$.

The aim of this paper which extends and completes previous studies of related configurations \cite{dec,alonso,rodluciano},
is to carefully reconsider cosmological models in which CDM and DE combine to behave as a gCg, modeling the dark sector of the Universe.
We shall consider the cosmic substratum as built of this dark sector together
with baryons and radiation.
Starting point of the numerical part is a confrontation of the background dynamics with the JLA sample of supernovae of type Ia \cite{jla}.
The parameter $\alpha$ which represents a measure of the distance to the $\Lambda$CDM model is poorly constrained by the SNIa data. However, the analysis of the JLA sample provides us with a range of values for the present dark matter fraction $\Omega_{c0}$ for any admissible $\alpha$. This information is then used to calculate the CMB and the matter power spectra
with the help of the CLASS code \cite{class}.
The observed CMB spectrum puts strong limits on $\alpha$ which is restricted to values very close to the $\Lambda$CDM limit $\alpha =0$. These limits are consistent with those obtained by a comparison of the gCg based matter power spectrum with its $\Lambda$CDM counterpart.

The structure of the paper is as follows.
In section \ref{gcg} we recall basic relations for the gCg and introduce its  decomposition
into an interacting  system of nonrelativistic matter and a DE component with constant EoS parameter.
On this basis we establish, in section \ref{model}, a cosmological four-component model by adding baryons and radiation. In section \ref{sn1a} the background dynamics is confronted with SNIa data from the JLA sample.
This analysis is also used in  section \ref{approx} to test the validity of an approximate analytic solution for the Hubble rate of the four-component system.
Section \ref{perturbations} is devoted to the system of first-order perturbation equations and provides the basis for the application of the CLASS code in  section \ref{numerical}.
Our results are summarized in section \ref{summary}.

\section{Generalized Chaplygin gas}
\label{gcg}

We start by modeling the cosmic medium as a one-component gCg with
a variable EoS parameter (cf. (\ref{chaplygin}))
\be
w = \frac{p}{\rho}=-\dfrac{A}{\rho^{\alpha+1}}, \qquad (-1\leq w<0),
\n{gcgeos}
\ee
which enters the energy conservation equation,
\be
\dot{\rho}+3H\left(1+w\right)\rho=0.
\n{gcgcontinuity}
\ee
A dot denotes the derivative with respect to cosmic time, $a$ is the scale factor of the Robertson-Walker metric and
$H=\frac{\dot{a}}{a}$ is the Hubble parameter.
The solution of the continuity equation (\ref{gcgcontinuity}) is
\be
\rho=\left[A+\dfrac{\left(1-A\right)}{a^{-3\left(1+\alpha\right)}}\right]^{\frac{1}{1+\alpha}}.
\n{rhogcg}
\ee
The present value of the scale factor was taken to be $a_{0} =1$. This solution represents a unification of the dark
sector in the sense that it behaves as matter, $\rho \propto a^{-3}$ for $a\ll 1$ and $\rho \approx $ constant for $a\gg 1$.

Considering a spatially flat universe, the system dynamics is given by
Friedmann's equation
\begin{equation}\label{}
3H^{2}=8\pi G\rho \n{friedmann1}
\end{equation}
and by
\ben
&\dot{H}=- 4\pi G\left(\rho+p\right).\n{friedmann2}
\een
The gCg is now split into a pressureless component, denoted by a subindex $c$, which is identified with CDM,  and a remaining part, denoted by a subindex $\Lambda$, which is supposed to represent a form of
DE, characterized by an EoS  $p_{\Lambda}=w_{\Lambda}\rho_{\Lambda}$ with a generally time-varying EoS parameter $w_{\Lambda}$,
\begin{equation}\label{}
\rho = \rho_{c} + \rho_{\Lambda}, \quad p = p_{c} + p_{\Lambda} = p_{\Lambda} = w_{\Lambda}\rho_{\Lambda}.
\end{equation}
The total pressure of the fluid is due the DE pressure.
For a semi-analytic treatment an explicit dependence $w_{\Lambda}(a)$ is required.
We shall restrict ourselves in the following to a constant $w_{\Lambda}$. Later on we shall focus on the
case $w_{\Lambda}=-1$ which is usually associated with a time-varying vacuum energy.
Then,
using the Friedmann equation (\ref{friedmann1}), one has
\be
\rho_{\Lambda}= \frac{w}{w_{\Lambda}}\frac{3H^{2}}{8\pi G} =
\rho_{\Lambda 0}\left(\dfrac{H}{H_{0}}\right)^{-2\alpha}.
\n{degcg}
\ee
For the special case $w_{\Lambda}=-1$ we recover the corresponding relation of \cite{nadgcg}.
With $w_{\Lambda}=-1$ and $\alpha=0$ the $\Lambda$CDM model is reproduced. The decaying vacuum model of \cite{Borges}
corresponds to $w_{\Lambda}=-1$ and $\alpha=-\frac{1}{2}$.

The separation of the gCg into two components is accompanied by an interaction between them.
With (\ref{degcg}) and assuming
\ben
&\dot{\rho}_{\Lambda}+3H\left(1+w_{\Lambda}\right)\rho_{\Lambda}=Q, \n{debalance}\\
&\dot{\rho}_{c}+3H\rho_{c}=-Q,
\n{cdmbalance}
\een
the source (loss) term $Q$ is found to be
\be
Q=3H\left[\alpha\left(1+w\right)+1+w_{\Lambda}\right]\rho_{\Lambda}.
\n{qint}
\ee
Notice that the term in the square brackets is not constant. It approaches a constant value only
in the high-redshift limit  $w \ll 1$. It is only in this limit that the frequently used
dependence $Q \propto H\rho_{\Lambda}$ (see, e.g. \cite{verde,salvatelli,clemson,wangAbd1}) is approximately valid.
If $Q>0$ the CDM component decays into DE, if $Q<0$ the
DE component decays into CDM.
The sign is determined by the interplay between $\alpha$ and $w_{\Lambda}$ in the square bracket of (\ref{qint}).
Since $w\geq -1$, the direction of the energy flux is defined by the sign of $\alpha$ for $w_{\Lambda}=-1$.
For $\alpha=0$ and
$w_{\Lambda}=-1$ the interaction vanishes and we consistently  recover the $\Lambda$CDM model.
Note that if only $\alpha=0$ we do not
recover the $w$CDM model, but we have a coupling that
is proportional to $1+w_{\Lambda}$.

The interaction term (\ref{qint}) may be rewritten as
\be
Q=3H\left[\alpha\frac{\rho_{c}\rho_{\Lambda}}{\rho} + \left(1+w_{\Lambda}\right)\left(\alpha\frac{\rho_{\Lambda}}{\rho}+1\right)\rho_{\Lambda}\right].
\n{qint2}
\ee
For $w_{\Lambda}=-1$ the second term in the square
bracket does not contribute. For this special case the
interaction assumes a nonlinear structure similar
to the cases studied in \cite{alonso,zhuk,FengZhang,rodluciano}.
In the following section we extend this simplified model to include baryons and radiation.

\section{gCg based cosmological model}
\label{model}

From now on we consider a universe composed of four components
described as perfect fluids: radiation (subindex $r$),
baryonic matter (subindex $b$),
CDM, (subindex $c$) and DE (subindex $\Lambda$),
\begin{equation}\label{}
\rho = \rho_{c} + \rho_{\Lambda}+ \rho_{r}+ \rho_{b} , \quad p = p_{\Lambda} + p_{r}.
\end{equation}
Because of the radiation component equation (\ref{degcg})
is no longer exactly valid in our four-component system.
But we continue to use it as an \textit{ansatz}.
Radiation and baryonic matter will be treated as separately conserved components.

Our model faces the problem that in the presence of radiation
the Hubble rate is no longer analytically known.
Introducing the dimensionless quantity $E$ by
\be
H\left(a\right)\equiv H_{0}E\left(a\right),
\n{E}
\ee
as well as
\begin{equation}\label{}
\Omega_{\Lambda 0} = \frac{8\pi G\rho_{\Lambda 0}}{3H_{0}^{2}}, \quad \Omega_{r0}
= \frac{8\pi G\rho_{r 0}}{3H_{0}^{2}},
\end{equation}
use of (\ref{friedmann1}), (\ref{friedmann2}) and  (\ref{degcg}) provides us with the
following differential equation for the Hubble rate:
\be
2\dfrac{dE}{da}Ea=-3E^{2}-3w_{\Lambda}\Omega_{\Lambda 0}E^{-2\alpha}-\Omega_{r0}a^{-4}.
\n{diffh}
\ee
For a negligible radiation component
equation (\ref{diffh}) has the analytical solution
\be
H=H_{0}\left\lbrace -w_{\Lambda}\left(1-\Omega_{m0}\right)
+\left[1+w_{\Lambda}\left(1-\Omega_{m0}\right)\right]
a^{-3\left(1+\alpha\right)}\right\rbrace^{\frac{1}{2\left(1+\alpha\right)}},\quad (\Omega_{r0} = 0),
\n{h}
\ee
where
\begin{equation}\label{}
\Omega_{c 0} = \frac{8\pi G\rho_{c 0}}{3H_{0}^{2}}, \quad \Omega_{b0}
= \frac{8\pi G\rho_{b 0}}{3H_{0}^{2}}, \quad \Omega_{m0}=\Omega_{c0}+\Omega_{b0}.
\end{equation}
Such solution, for $\alpha = -0.5$, has been used for a SNe Ia analysis in \cite{sndec}).
However, at high redshift the radiation component plays an important
role. To take into account the radiation component properly, we use the analytical approximation
\be
H=H_{0}\sqrt{\left[-w_{\Lambda}\left(1-\Omega_{m0}\right)+
\left[1+w_{\Lambda}\left(1-\Omega_{m0}\right)\right]
a^{-3\left(1+\alpha\right)}\right]^{\frac{1}{1+\alpha}}+\Omega_{r0}a^{-4}}.
\n{happrox}
\ee
This expression is obtained by adding a radiation contribution in
 (\ref{h}).
The viability of this approximation will be tested below.
Note that in equation (\ref{happrox}) the terms $\Omega_{m0}$ and
$w_{\Lambda}$ only appear in the combination $w_{\Lambda}\left(1-\Omega_{m0}\right)$.
This means, $\Omega_{m0}$ and
$w_{\Lambda}$ are not separate degrees of freedom here.
Therefore it is useful to define a variable $\tilde{\Omega}_{m0}$ by
\be
1-\tilde{\Omega}_{m0}=-w_{\Lambda}\left(1-\Omega_{m0}\right).
\n{omega}
\ee
This degeneracy, which is not a consequence of our approximation, means that any cosmological test which relies on the Hubble rate cannot constrain $w_{\Lambda}$ and $\Omega_{c0}$ (or equivalently $\Omega_{\Lambda 0}$) at the same time.
For $w_{\Lambda}=-1$ the parameters $\tilde{\Omega}_{m0}$ and $\Omega_{m0}$ coincide.
In terms of $\tilde{\Omega}_{m0}$ the Hubble rate (\ref{happrox}) can then
be written as
\be
H=H_{0}\sqrt{\left(1-\tilde{\Omega}_{m0}+\tilde{\Omega}_{m0}a^{-3\left(1+\alpha\right)}\right)^{\frac{1}{1+\alpha}}+\Omega_{r0}a^{-4}}.
\n{happrox1}
\ee
Using this approximate solution in (\ref{degcg}), the DE energy density becomes
\be
\rho_{\Lambda}=\frac{3H^{2}_{0}}{8\pi G}\Omega_{\Lambda 0}\left[\left(1-\tilde{\Omega}_{m0}+\tilde{\Omega}_{m0}a^{-3\left(1+\alpha\right)}\right)^{\frac{1}{1+\alpha}}+\Omega_{r0}a^{-4}\right]^{-\alpha},
\n{rhol}
\ee
where $\Omega_{\Lambda 0}=1-\Omega_{m0} -\Omega_{r0}$.
The CDM energy density is found through
\be
\rho_{c}=\frac{3H^{2}}{8\pi G}-\rho_{\Lambda}-\rho_{b}-\rho_{r},
\n{rhocdm}
\ee
where $\rho_{b}$ and $\rho_{r}$ are given by
\ben	
&\rho_{b}=\dfrac{3H^{2}_{0}}{8\pi G}\Omega_{b0}a^{-3},
\een
and
\ben
&\rho_{r}=\dfrac{3H^{2}_{0}}{8\pi G}\Omega_{r0}a^{-4},
\een
respectively.
In the following section we confront this background dynamics with the binned SNIa data from the
JLA sample \cite{jla}.
Since even the most distant supernovae have a low redshift
(compared with the redshift of the last-scattering surface), the radiation component in the energy
balance is small and the approximate solution is justified for this analysis.

\section{Supernovae statistical analysis}
\label{sn1a}

The baryon-photon subsystem of the cosmic medium will be treated here in the same manner as it is treated in the $\Lambda$CDM model.
We fix $\Omega_{b0}$ and $\Omega_{r0}$ according to their  best-fit values
in \cite{planck}.
We divide our analysis into two parts. At first we consider the dynamics for $w_{\Lambda}=-1$
with the three free parameters $\Omega_{c0}$, $h$ and $\alpha$, where $h$ is introduced as usual by
$H_{0}=100h\ \mathrm{kms^{-1}Mpc^{-1}}$.
In the second part we deal with the general case $w_{\Lambda}\neq -1$ with the free parameters
$\tilde{\Omega}_{m0}$, $h$ and $\alpha$.
In the general case we use equation
(\ref{happrox1}) for the Hubble rate, while for $w_{\Lambda}=-1$
we have the explicit expression
\be
H=H_{0}\sqrt{\left[1-\left(\Omega_{c0}+\Omega_{b0}\right)+\left(\Omega_{c0}+\Omega_{b0}\right)
a^{-3\left(1+\alpha\right)}\right]^{\frac{1}{1+\alpha}}+\Omega_{r0}a^{-4}}\qquad (w_{\Lambda}=-1).
\n{happrox2}
\ee
%ote that mathematically the both cases only differ by a translation,
%but they have different physical meanings.
The results for  the time varying vacuum model ($w_{\Lambda}=-1$) will be used to establish a
$2\sigma$ range of admissible values for each of the free parameters. Posteriorly, in order to
obtain the CMB power spectrum, we use
this range as a \textit{prior} to compare the approximate solution of the
Hubble rate (\ref{happrox2}) against the numerical solution of the
differential equation (\ref{diffh}) with $w_{\Lambda}=-1$.
A similar analysis for a general EoS will be used to constrain $\Omega_{c0}$ and
$w_{\Lambda}$.

As is well known, SNe Ia tests are using the luminosity distance
modulus (the superscript ``th" means ``theoretical"),
\be
\mu^{th}=5\log\left[d_{L}\left(z\right)\right]+\mu_{0},
\n{muSN}
\ee
with $\mu_{0}=42.384-5\log\left(h\right)$, where
\be
d_{L}\left(z\right)=\left(z+1\right)H_{0}\int_{0}^{z}\dfrac{d\tilde{z}}{H\left(\tilde{z}\right)}
\ee
is the luminosity distance.
The crucial quantity for the statistical analysis is
\be
\chi^{2}_{SN}=\Delta\mu^{T}\left(\theta\right)\ \mathcal{C}\ \Delta\mu\left(\theta\right).
\ee
Here, $\mathcal{C}$ denotes the covariance matrix and
$\Delta\mu$ is a vector whose i-th component is given by
$\Delta\mu_{i}=\mu^{obs}_{i}-\mu^{th}\left(z_{i}\right)$, where the observational distance modulus
$\mu^{obs}$ has the structure \cite{jla}
\be
\mu^{obs}=m_{M}^{*}-\left(M_{B}-\alpha X_{1}+\beta C\right).
\ee
The quantities $\alpha$ (here \textit{not} the gCg parameter), $\beta$ and $M_{B}$ are nuisance parameters and
$m_{M}^{*}$, $X_{1}$ and $C$ are light-curve parameters.
Since the model has an isotropic
luminosity distance, it is possible to use the 31 binned data and the
corresponding covariance matrix of \cite{jla}.
In TABLE~\ref{tabbg} we present the $2\sigma$ confidence level
constraints of this analysis for the case
$w_{\Lambda}=-1$.
The resulting distance modulus for the best-fit values of TABLE~\ref{tabbg} is visualized and compared with the $\Lambda$CDM values in FIG.~\ref{figmu}.
The scale-factor dependences of the fractional abundances of the components and of the Hubble rate are displayed by FIG.~\ref{figomegaref}. For comparison we include also the $\Lambda$CDM results. One expects that differences in those background quantities which appear as coefficients in the perturbation equations will also affect the CMB spectrum.

FIG.~\ref{figbg} shows the marginalized $1\sigma$ and $2\sigma$ contours and the
probability distribution functions (PDFs) for each of the
free parameters.
 Note that the SNe Ia
analysis does not impose strong constraints on the gCg parameter $\alpha$.
Both $\alpha =0$ and $\alpha =-1$ are within the $2\sigma$ confidence region.
\begin{table}[h!]
\centering
\caption{Result of SNe Ia statistical analysis for the time varying vacuum model ($w_{\Lambda}=-1$).}
\centering
\begin{tabular}{ccc}
\hline\hline
$\qquad\Omega_{c0}\qquad$ & $\qquad h\qquad$ & $\qquad\alpha\qquad$\\ \hline\hline
$\qquad 0.317\pm_{0.144}^{0.078} \qquad$ & $\qquad 0.697\pm_{0.012}^{0.019} \qquad$ & $\qquad -0.528\pm_{0.540}^{0.729} \qquad$ \\\hline
\end{tabular}
\n{tabbg}
\end{table}

\begin{figure}[h!]
\includegraphics[scale=0.8]{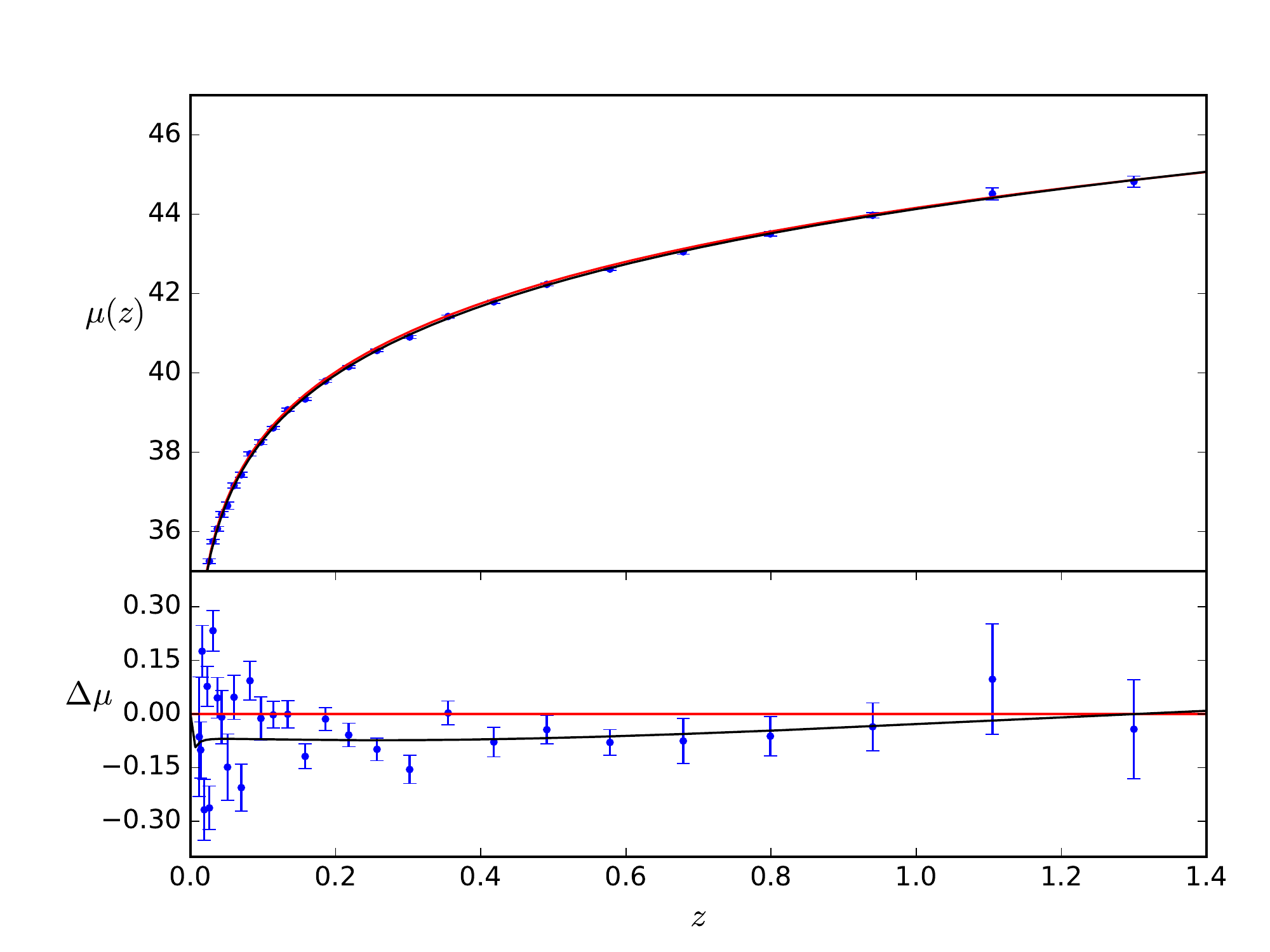}
\caption{Upper part: distance modulus calculated from Eq.~(\ref{muSN}) with the best-fit values of TABLE~\ref{tabbg} (black curve). The red curve represents the $\Lambda$CDM model  with the best-fit values of \cite{jla}.
Lower part: difference $\Delta\mu$ between the distance modulus of our model and the distance modulus of the $\Lambda$CDM model.}
\n{figmu}
\end{figure}

\begin{figure}[h!]
\includegraphics[scale=0.8]{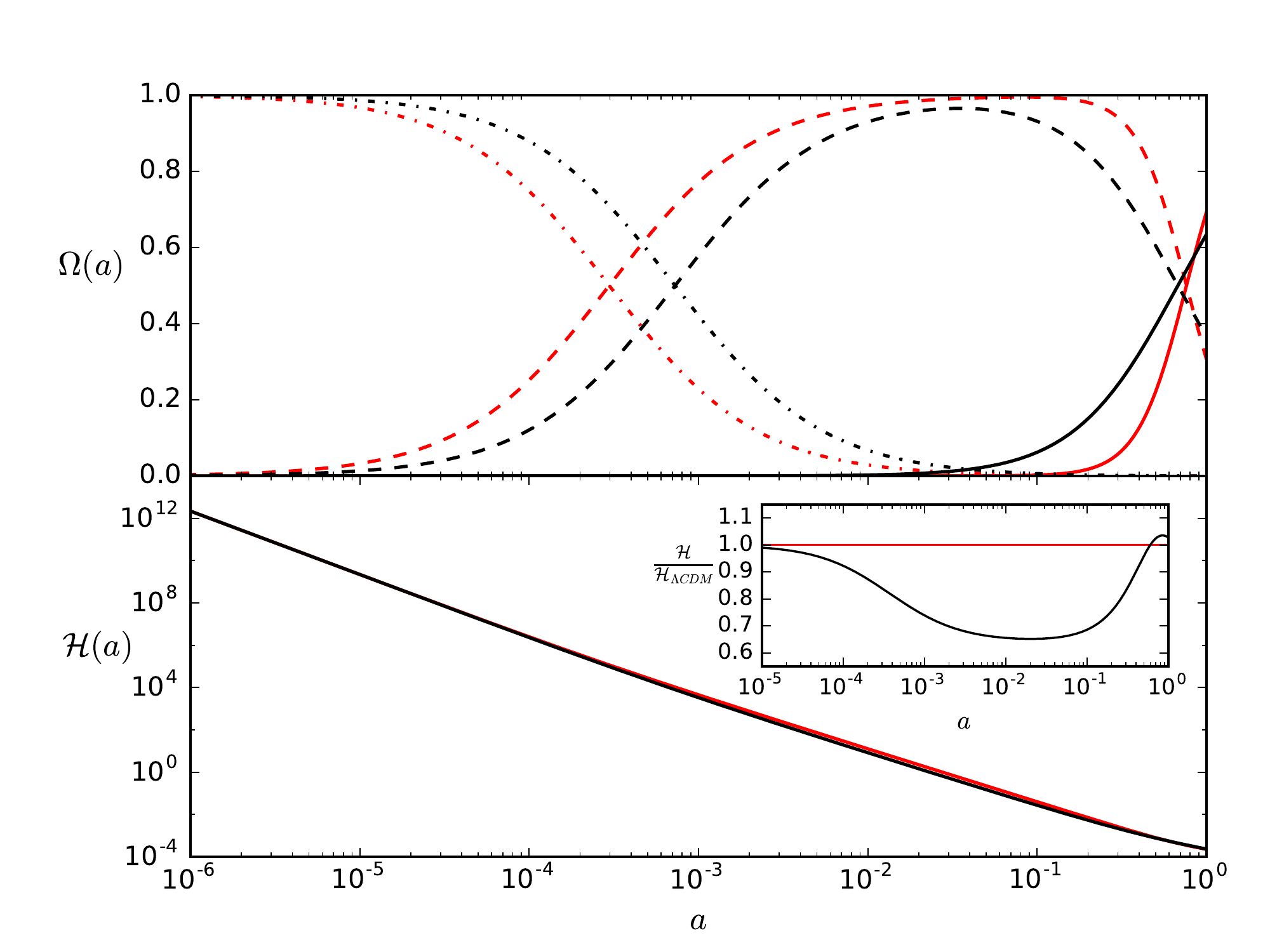}
\caption{Upper part: fractional abundances of the components. The dot-dashed lines describe the radiation fractions, the dashed lines the matter fractions and the solid lines the DE fractions (black for our model, red for $\Lambda$CDM).  Lower part: Hubble rate for the present model compared with that for the $\Lambda$CDM model.}
\n{figomegaref}
\end{figure}

\begin{figure}[h!]
\begin{flushleft}
\subfigure{
\includegraphics[width=5cm,height=5cm,keepaspectratio]{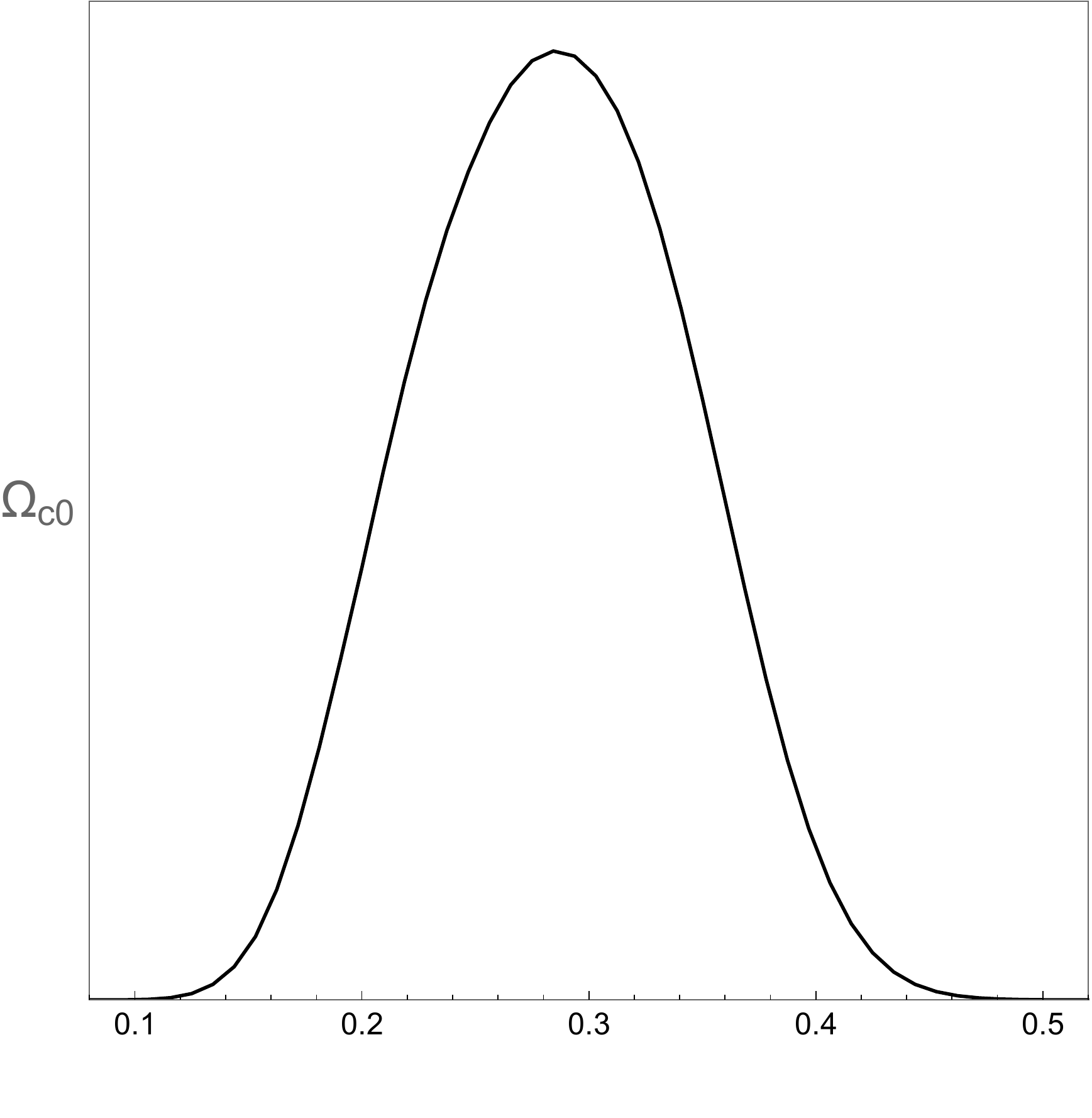}}\\
\end{flushleft}
\begin{flushleft}
\subfigure{
\includegraphics[width=5cm,height=5cm,keepaspectratio]{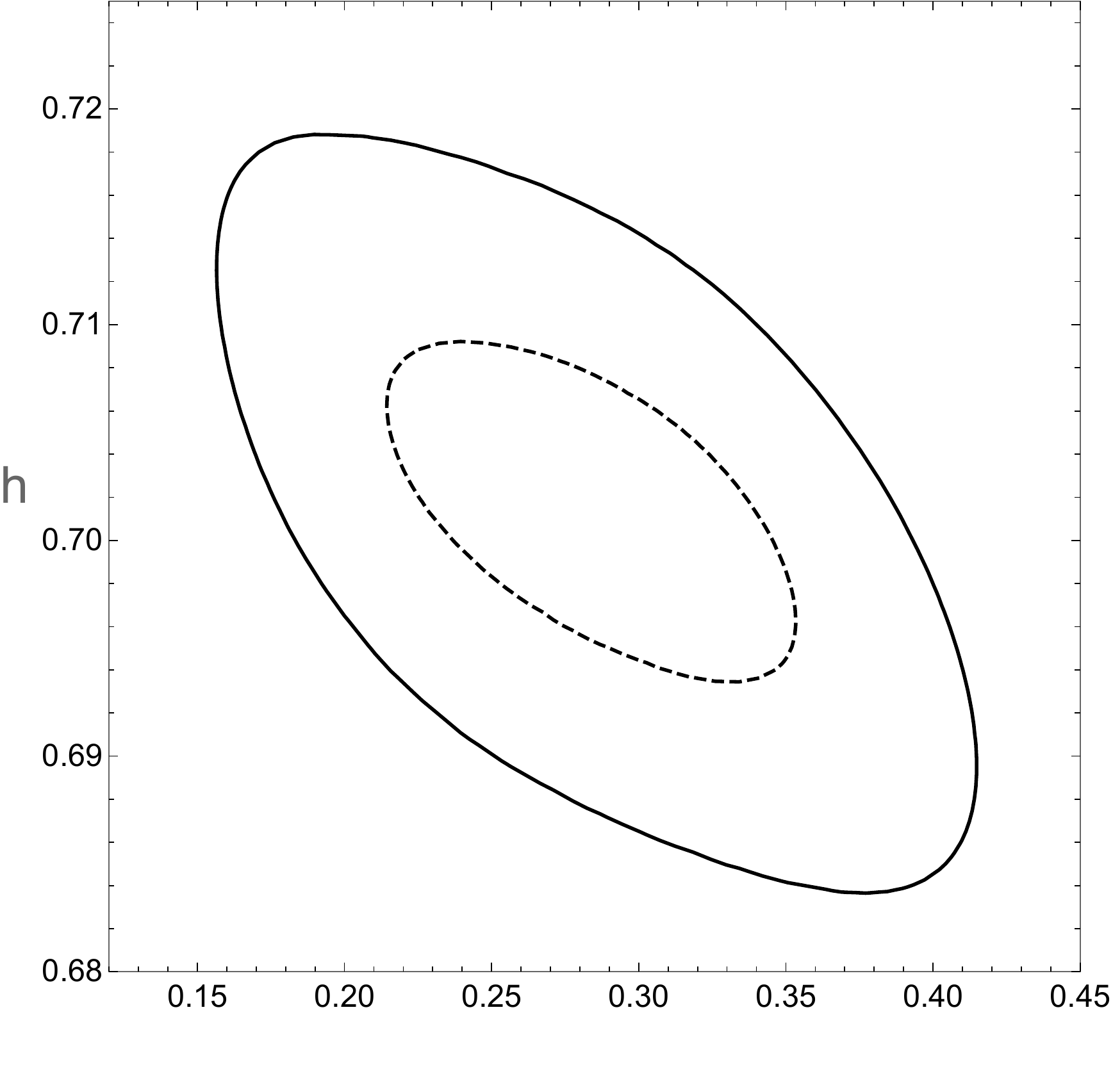}}
\subfigure{
\includegraphics[width=5cm,height=5cm,keepaspectratio]{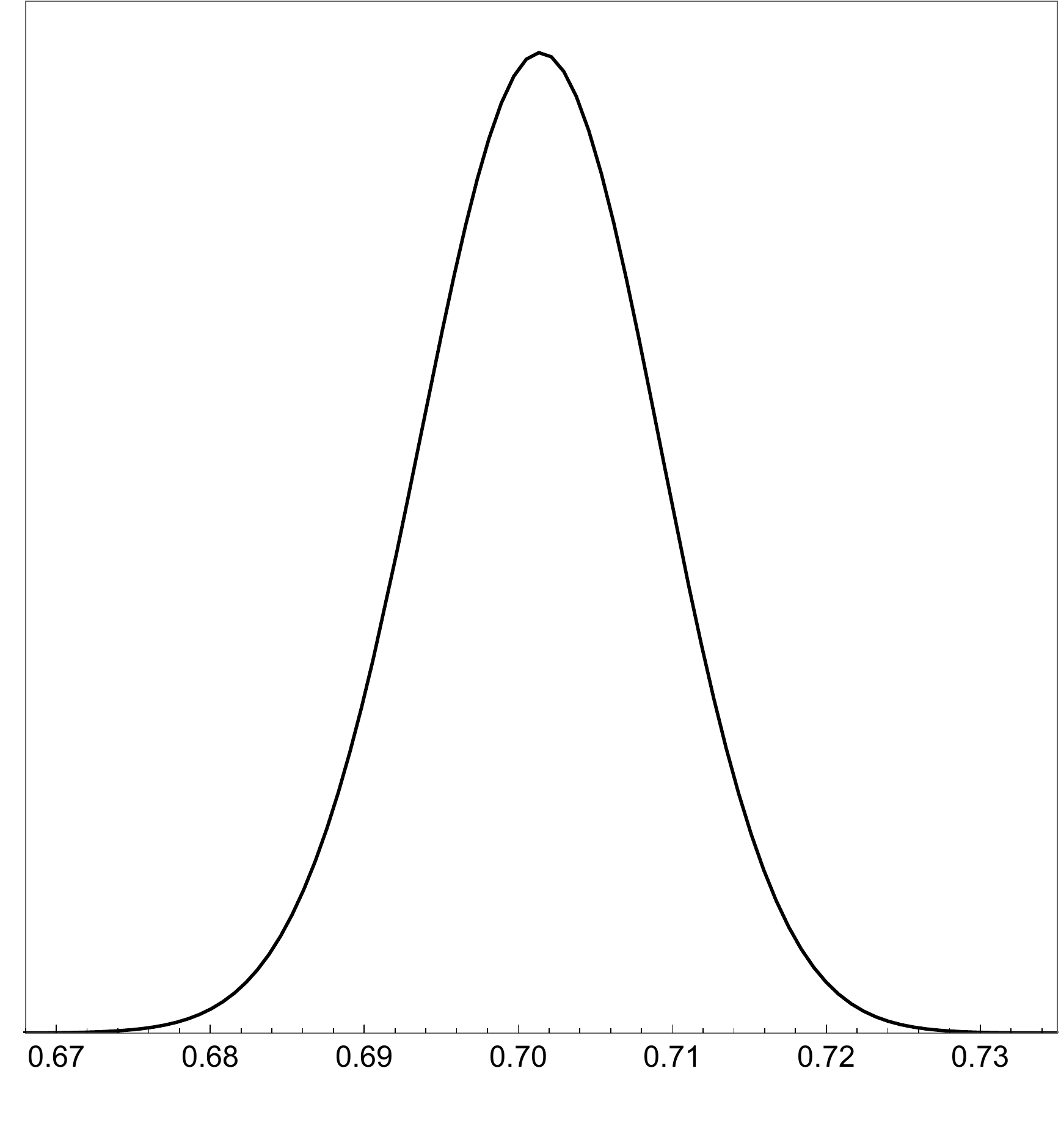}}\\
\end{flushleft}
\begin{flushleft}
\subfigure{
\includegraphics[width=5cm,height=5cm,keepaspectratio]{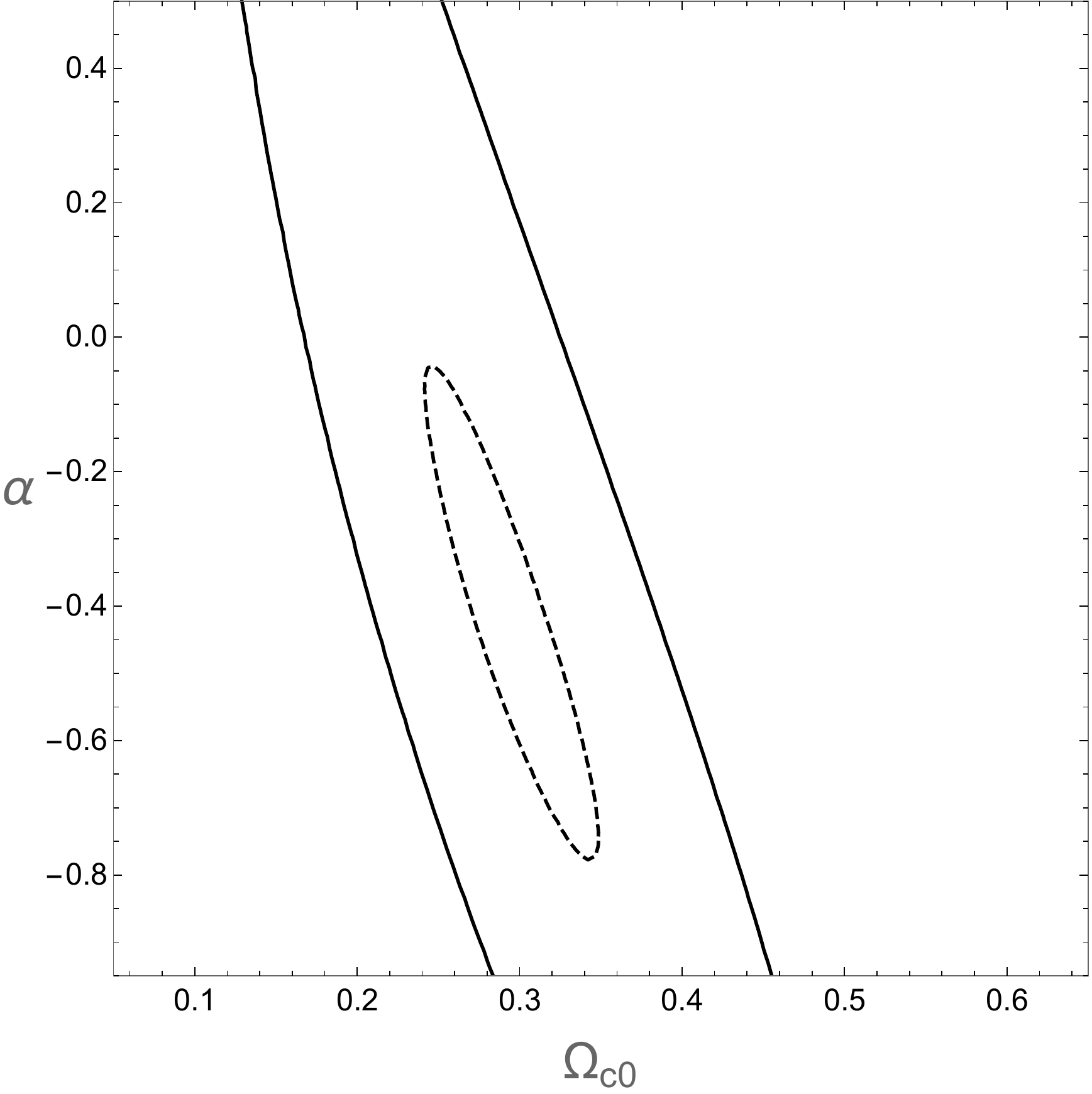}}
\subfigure{
\includegraphics[width=5cm,height=5cm,keepaspectratio]{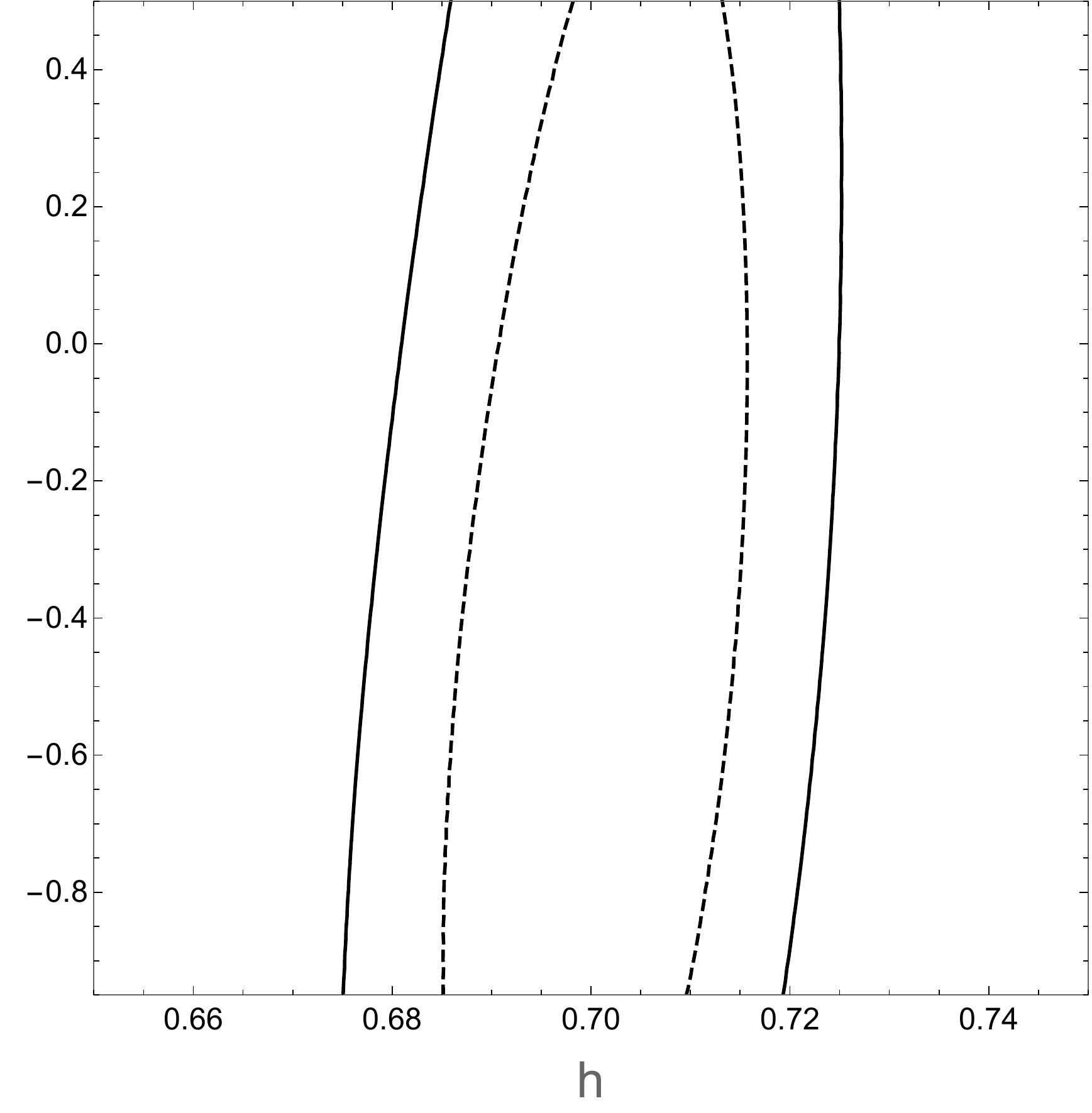}}
\subfigure{
\includegraphics[width=5cm,height=5cm,keepaspectratio]{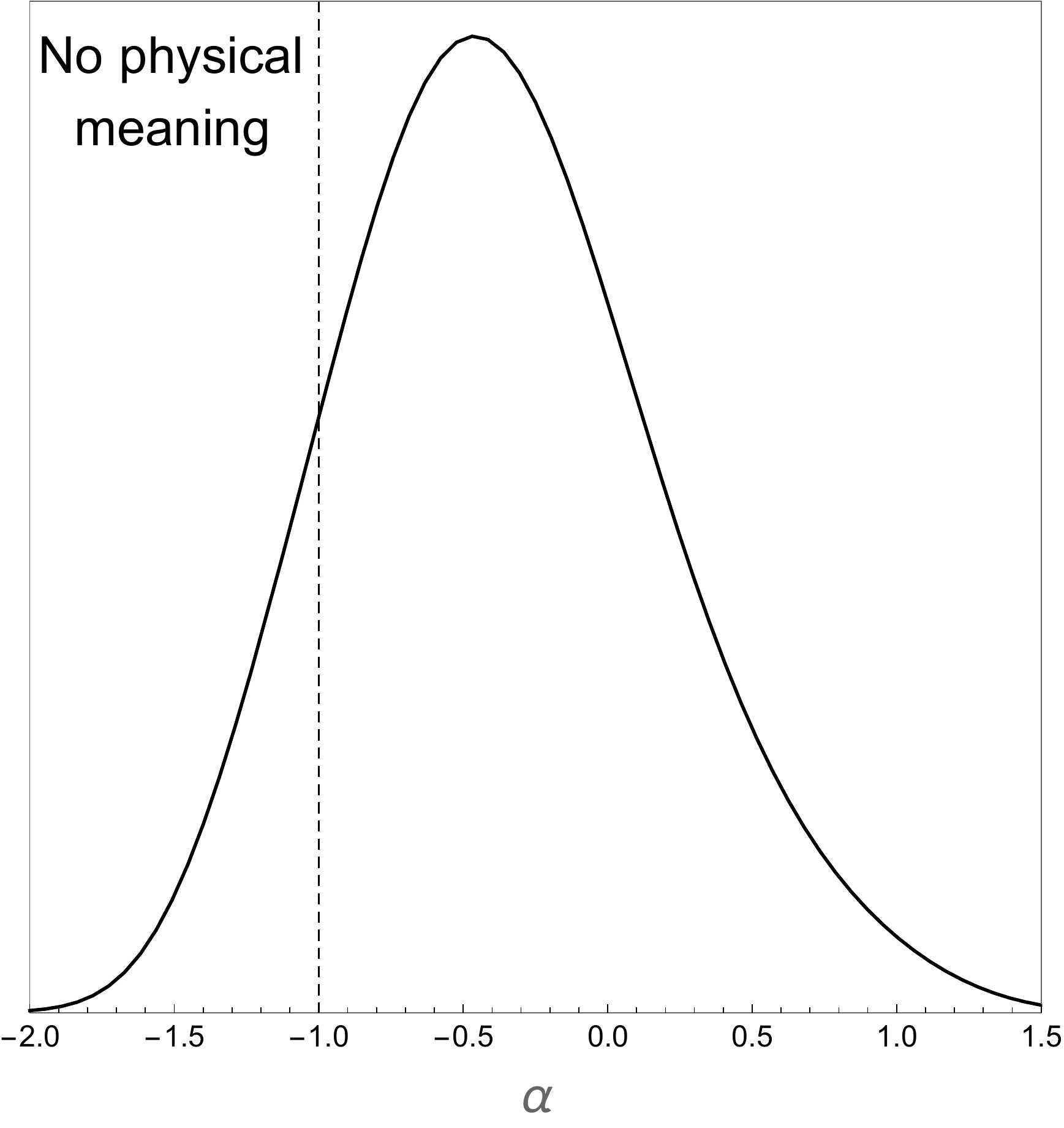}}\\
\end{flushleft}
\caption{Marginalized results of SNe Ia statistical analysis for the
time-varying vacuum model. Adapting the baryon fractions,
this result coincides with the results presented in \cite{saulo14,3page}.}
\n{figbg}
\end{figure}

For the general case the SNe Ia statistical analysis
provides us with $\tilde{\Omega}_{m0}=0.363\pm_{0.145}^{0.075}$. Using
this result together with equation (\ref{omega}) we may infer a range of admissible combinations
of $\Omega_{m0}$ and $w_{\Lambda}$. FIG.~\ref{figomega} shows
the allowed region in the $w_{\Lambda}-\Omega_{m0}$ plane.
\begin{figure}[h!]
\includegraphics[scale=1]{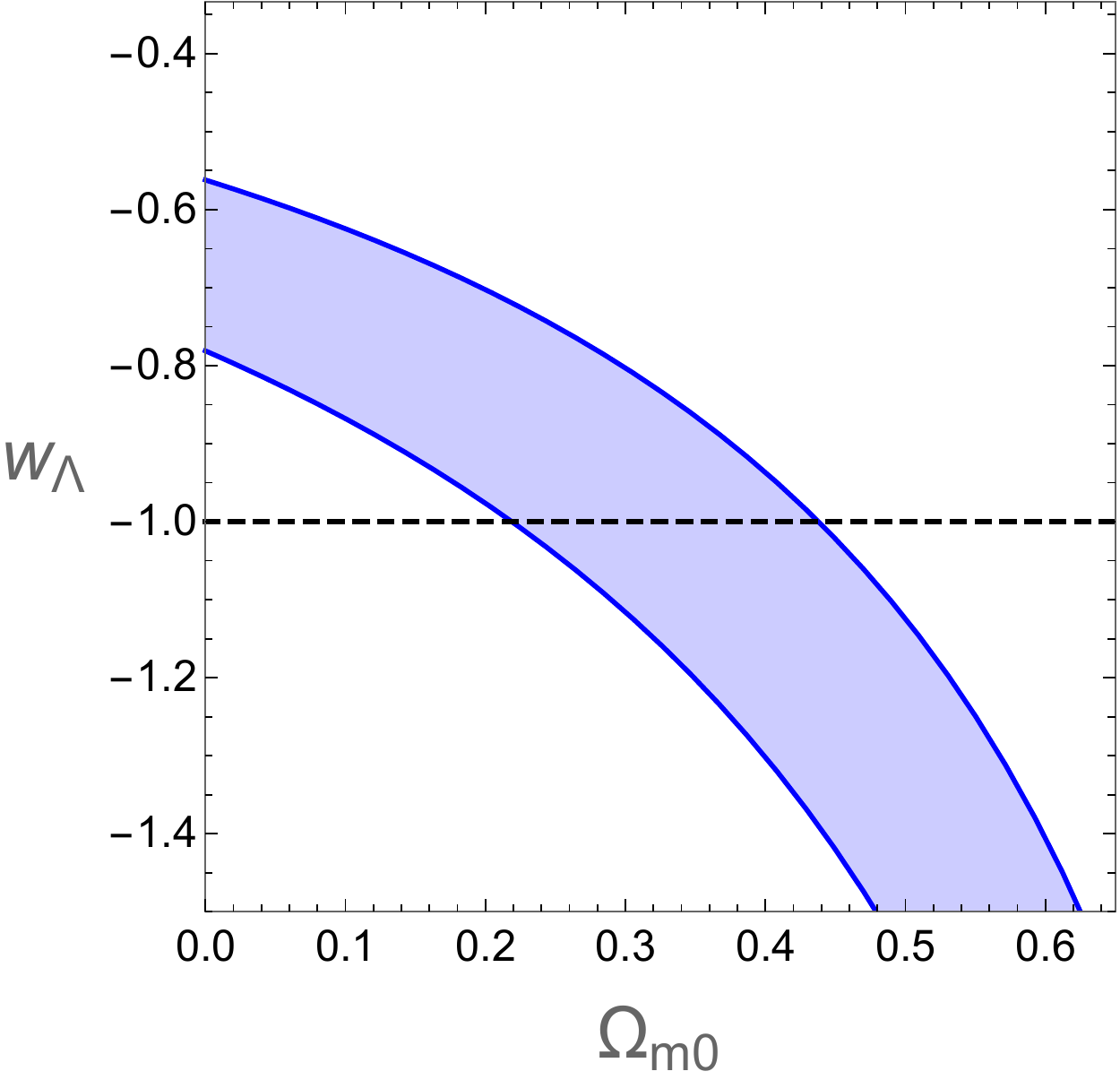}
\caption{Region in the $w_{\Lambda}-\Omega_{m0}$ plane that satisfies the
constraint (\ref{omega}), using the $2\sigma$ limits for
$\tilde{\Omega}_{m0}$ of the SNe Ia statistical analysis.}
\n{figomega}
\end{figure}
Since $\Omega_{c0}$ and $w_{\Lambda}$ are only weakly constrained separately by this analysis, we focus
on the vacuum model $w_{\Lambda}=-1$ in the following.

\section{Approximate solution for the Hubble rate}
\label{approx}

For the SNIa analysis of the previous section the r\^{o}le of the radiation component was marginal.
For a study of the CMB anisotropy spectrum its appropriate inclusion is essential, however.
This raises the question of whether the expression  (\ref{happrox2}) is a good approximation to the exact solution
of the differential equation (\ref{diffh}) for $w_{\Lambda}=-1$.
Since we prefer to work with an analytic solution we shall test its viability by comparing it with the numerical solution of (\ref{diffh}).
To this purpose we define the deviation between the approximate analytic and the numerical solutions by
\be
\mbox{Deviation}=\dfrac{H_{\mbox{num.}}-H_{\mbox{approx.}}}{H_{\mbox{num.}}}.
\n{deviation}
\ee
Note that, since the dependence on $h$ appears only in $H_{0}$,
and the Hubble rate can be written as (\ref{E}), this deviation depends
only on $\Omega_{c0}$ and $\alpha$.
Our interest is to find a range in which we can use the
approximate solution (\ref{happrox2}) in the high-redshift regime. Given
that the studied model can be seen as a generalization of the $\Lambda$CDM
model and the parameter that measures the difference to the latter is $\alpha$,
we expect the deviation to be very sensitive to variations in $\alpha$.
In order to calculate the deviation we proceed in the following way:
first we fix a parameter $\alpha$ within the
$2\sigma$ interval that was obtained in the SNe Ia analysis.
Then we calculate the deviations (\ref{deviation}) by using separately the upper and lower limits of the
$2\sigma$ range for $\Omega_{c0}$ (i.e., $0.317- 0.144$ and  $0.317 + 0.078$),  found by the SNIa analysis.
This results in two curves for (\ref{deviation}) in dependence on the scale factor.
These curves which confine the error regions for the chosen $\alpha$ are shown in FIG.~\ref{figdeviation}
for several values of $\alpha$.
For all these cases the upper curves correspond to $\Omega_{c0} = 0.317- 0.144$, the lower ones to
$\Omega_{c0} = 0.317 + 0.078$.
Increasing $\alpha$  allows us to push the deviation to less than $1\%$.
 The results are also summarized in TABLE~\ref{tabdeviation}. Interestingly, the maximal error of the approximate solution does not occur at the highest redshift but at some intermediate value.
For positive values of $\alpha$ the deviation is negligible.

\begin{figure}[h!]
\includegraphics[scale=0.6]{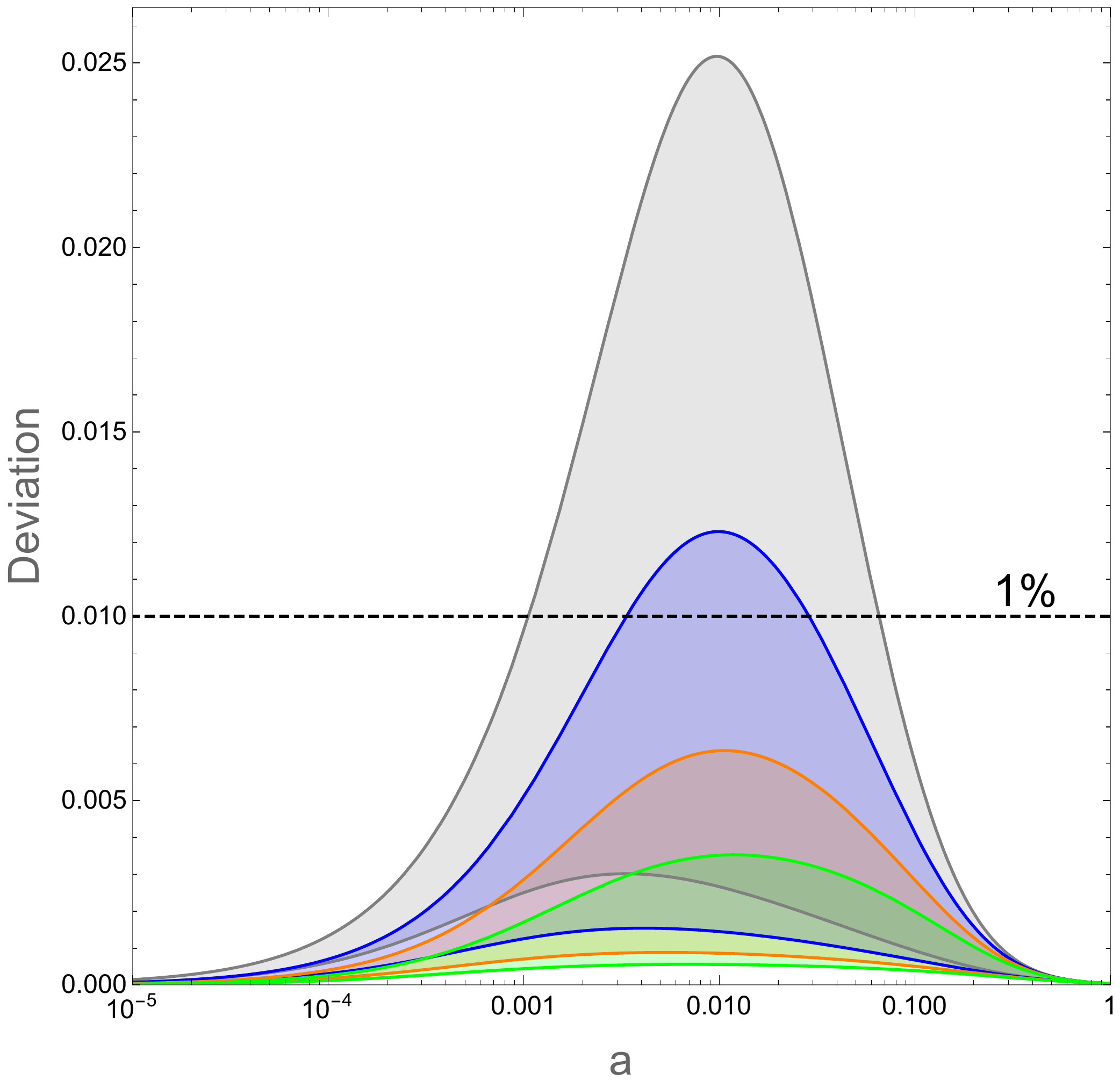}
\caption{Deviation between the approximate solution (\ref{happrox2}) and
the numerical solution of (\ref{diffh}).
Each colored region is associated with a certain value of $\alpha$.
(Gray for $\alpha=-0.65$, blue for $\alpha=-0.60$, orange for $\alpha=-0.55$ and green for $\alpha=-0.50$).
All regions are bound by a lower curve, corresponding to the upper $2\sigma$-limit $\Omega_{c0} = 0.317 + 0.078$ and an upper curve for the lower-$2\sigma$ limit
$\Omega_{c0} = 0.317- 0.144$.
}
\n{figdeviation}
\end{figure}
\begin{table}[h!]
\centering
\caption{DEVIATION (\ref{deviation}).}
\centering
\begin{tabular}{ccc}
\hline\hline
$\qquad\Omega_{c0}\qquad$&$\qquad\alpha\qquad$&$\qquad\mbox{Maximum deviation}\qquad$\\ \hline\hline
\multirow{5}{*}{$0.317\pm_{0.144}^{0.078}$}&$\qquad-0.8$&$\qquad 20,0\%$ \\
&$\qquad-0.7$&$\qquad 5,28\%$ \\
&$\qquad-0.6$&$\qquad 1,23\%$ \\
&$\qquad-0.5$&$\qquad 0.353\%$ \\
&$\qquad-0.4$&$\qquad 0.135\%$ \\
\hline
\end{tabular}
\n{tabdeviation}
\end{table}

In view of these results for the deviation (\ref{deviation}) of the approximate expression (\ref{happrox2}) from the exact solutions
we shall impose from now on the \textit{prior} $\alpha>-0.55$ in order
to keep the error well below the $1\%$ level.

\section{Perturbation dynamics}
\label{perturbations}

Restricting ourselves to scalar perturbations in a spatially flat universe, the perturbed Robertson-Walker metric
in the Newtonian gauge with the scalar degrees of freedom $\psi$ and
$\phi$ is
\be
ds^{2}=a^{2}\left(\tau\right)\left[-\left(1+2\psi\right)d\tau^{2}+\left(1-2\phi\right)dx^{i}dx_{i}\right].
\n{metric}
\ee
This metric leads to the set of  Einstein's equations \cite{ma&bertschinger}
\ben
&k^{2}\phi+3\mathcal{H}\left(\phi^{\prime}+\mathcal{H}\psi\right)=-4\pi Ga^{2}\hat{\rho},\\
&\phi^{\prime}+\mathcal{H}\psi=-4\pi Ga^{2}\hat{v},\\
&\phi^{\prime\prime}+\mathcal{H}\left(\psi^{\prime}+2\phi^{\prime}\right)+\left(2\mathcal{H}^{\prime}+\mathcal{H}^{2}\right)\psi+\dfrac{k^{2}}{3}\left(\phi-\psi\right)=\dfrac{4\pi}{3}Ga^{2}\hat{p},\\
&k^{2}\left(\phi-\psi\right)=12\pi Ga^{2}\left(\rho+p\right)\hat{\sigma}.
\een
Here, the prime denotes the derivative with respect to the conformal time $\tau$ and $\mathcal{H}\equiv \frac{a^{\prime}}{a}$ is the Hubble parameter with respect to the conformal time. The hat denotes first-order variables.
The quantity $\hat{\sigma}$ is associated with anisotropic
stress perturbations, $\hat{v}$ is the total peculiar velocity potential, related to the spatial components of the four-velocity by $\partial^{i}\hat{v}=a\ u^{i}$.

In order to obtain the CMB and linear matter power spectra we have to
solve the complete set of perturbation equations for all components of
the universe. The standard procedure to obtain the CMB temperature
anisotropies is to compute the Boltzmann equations for all these components.
Here we assume
that baryons and radiation behave in the same way as they do in the $\Lambda$CDM model, i.e.,
interacting with each
other via Thomson scattering before recombination but not directly with the dark sector.
Thus, the Boltzmann
equations for these two components will be the same as the well-established
equations of \cite{ma&bertschinger}.
However, since we do not have yet a microscopic description of the
interaction between the dark components, corresponding Boltzmann equations are not available either.
Instead,  we have to use the fluid dynamical description for the components of the dark
sector.

Generally, a cosmic fluid A with a perfect-fluid energy-momentum tensor
\begin{equation}
T^{\mu\nu}_{A}=\rho_{A}u^{\mu}_{A}u^{\nu}_{A}+p_{A}h^{\mu\nu}_{A},
\label{emtensor}
\end{equation}
obeys the energy-momentum balance
\begin{equation}
\nabla_{\mu}T^{\mu\nu}_{A}=Q^{\nu}_{A},
\label{cderivative}
\end{equation}
where  interactions with other components of the cosmic substratum are included by a source (loss) term
$Q^{\nu}_{A}$, which can be split according to
\begin{equation}
Q^{\mu}_{A}=Q_{A}u^{\mu}_{A}+\hat{F}^{\mu}_{A},\quad\quad\mbox{where}\quad\quad \hat{F}^{\mu}_{A}u_{A\mu}=0.
\label{qu}
\end{equation}
The crucial quantities for a
perturbative analysis are the density contrast $\delta_{A}$ and the peculiar
velocity potential $\hat{v}_{A}$, which are defined by,
\ben
&\delta_{A}=\dfrac{\hat{\rho}_{A}}{\rho_{A}},\\
&\partial^{i}\hat{v}_{A}=a\ u^{i}_{A},
\n{delta&u}
\een
where $u^{i}_{A}$ is the four-velocity of comonent $A$.
If the fluid has a constant EoS
parameter $w_{A}$, the energy balance takes the form
\ben
\delta^{\prime}_{A}&+&3\mathcal{H}\left(c^{2}_{sA}-w_{A}\right)\delta_{A}
\nonumber\\
&-&9\mathcal{H}^{2}\left(1+w_{A}\right)\left(c^{2}_{sA}-
w_{A}\right)\hat{v}_{A}-\left(1+w_{A}\right)\left(k^{2}\hat{v}_{A}+3\phi^{\prime}\right)
\nonumber\\
&&\qquad =a\dfrac{Q_{A}}{\rho_{A}}\left[\psi-\delta_{A}
-3\mathcal{H}\left(c^{2}_{sA}-w_{A}\right)\hat{v}_{A}\right]+a\dfrac{\hat{Q}_{A}}{\rho_{A}},
\label{kenergy}
\een
where $c^{2}_{sA}$ is the comoving sound speed and $\hat{Q}_{A}$ is the perturbation of the
temporal component of the interaction term.
Moreover, the general momentum balance is
\ben
\hat{v}^{\prime}_{A}+\mathcal{H}\left(1-3c^{2}_{sA}\right)\hat{v}_{A}
+\dfrac{c^{2}_{sA}}{1+w_{A}}\delta_{A}+\psi\qquad\qquad&&
\nonumber\\
=a\dfrac{Q_{A}}{\rho_{A}\left(1+w_{A}\right)}\left[\hat{v}
-\left(1+c^{2}_{sA}\right)\hat{v}_{A}\right]+\frac{a}{1+w_{A}}\dfrac{\hat{f}_{A}}
{\rho},&&
\label{kmomentum}
\een
where $\hat{f}_{A}$ 
is introduced through
\begin{equation}
\hat{F}^{i}_{A}=\dfrac{1}{a}\partial^{i}\hat{f}_{A}.
\end{equation}
For the case $w_{\Lambda}=-1$ the DE peculiar velocity potential
has no dynamics and the energy balance (\ref{kenergy}) reduces to,
\ben
&\delta^{\prime}_{\Lambda}+3\mathcal{H}\left(c^{2}_{s}+1\right)\delta_{\Lambda}
=a\dfrac{Q}{\rho_{\Lambda}}\left(\psi-\delta_{\Lambda}\right)+\dfrac{a}{\rho_{\Lambda}}\hat{Q},\qquad (w_{\Lambda}=-1),
\n{decdeltade}
\een
where we dropped the index $\Lambda$  in $Q$, i.e. $Q = Q_{\Lambda}$.
Since for $w_{\Lambda}=-1$ the DE peculiar velocity potential is not a dynamic variable,
we can
use equation (\ref{kmomentum}) to obtain the
spatial perturbation of the interaction term,
\be
\hat{f}=\dfrac{c_{s}^{2}\rho_{\Lambda}\delta_{\Lambda}}{a}-Q\hat{v},\qquad (w_{\Lambda}=-1).
\n{hatf}
\ee
For the CDM component the energy  and momentum balances are
\ben
&\delta_{c}^{\prime}-k^{2}\hat{v}_{c}-3\phi^{\prime}
=-a\dfrac{Q}{\rho_{c}}\left(\psi-\delta_{c}\right)+a\dfrac{\hat{Q}}{\rho_{c}},\qquad (w_{\Lambda}=-1),
\n{pertenergycdm}
\een
and
\ben
&\hat{v}^{\prime}_{c}+\mathcal{H}\hat{v}_{c}+\psi
=-a\dfrac{Q}{\rho_{c}}\left(\hat{v}-\hat{v}_{c}\right)-a\dfrac{\hat{f}}{\rho_{c}},\qquad (w_{\Lambda}=-1),
\n{pertmomentumcdm}
\een
respectively.
Note that in the equations above $Q$ is given by equation (\ref{qint}),
and $\hat{f}$ is given by equation (\ref{hatf}).

The perturbation $\hat{Q}$ of the interaction term  has to be chosen on physical grounds.
We assume that the expression (\ref{degcg}) continues to be valid at first order.
One realizes that (\ref{degcg}) can covariantly be written as $\rho_{\Lambda} = \rho_{\Lambda}\left(\frac{\Theta}{\Theta_{0}}\right)^{-2\alpha}$, where the expansion scalar  $\Theta \equiv u^{\mu}_{;\mu}$ reduces to $\Theta =3H$ in the background.
 Recall that in our four-component model (\ref{degcg}) is an ansatz,  motivated by the fact that it is an exact relation  in a two-component universe of CDM and DE which in total behaves as a generalized Chaplygin gas.
 This
assumption leads to a first-order DE density contrast
\be
\delta_{\Lambda}=-\dfrac{2\alpha}{3H}\hat{\Theta},
\n{pertdeltade}
\ee
where $\hat{\Theta}$ in the Newtonian gauge is
\be
\hat{\Theta}=\dfrac{1}{a}\left(\psi^{\prime}+\phi^{\prime}-k^{2}\hat{v}\right).
\n{hattheta}
\ee
The first-order source term is then obtained by introducing (\ref{pertdeltade}) in (\ref{decdeltade}) and
solving for $\hat{Q}$.

\section{Numerical computation}
\label{numerical}

Now we apply the CLASS code to the set of perturbation equations of the previous section.
Since we are looking for a concordance model, we use the $2\sigma$ range of $\alpha$ values,
obtained from the SNe Ia statistical analysis in section \ref{sn1a}.
The procedure is as follows:  we fix an $\alpha$ value out of the $2\sigma$ confidence interval of the
SNIa analysis.
Then we compute the
CMB and matter power spectra using the upper and the lower limits of the $2\sigma$ confidence interval for
$\Omega_{c0}$.  Since these limits depend on $\alpha$ (see the $\Omega_{c0} - \alpha$ contour curves in FIG.~\ref{figbg}),
we have to choose a slightly different range of
$\Omega_{c0}$ values for each value of $\alpha$. TABLE~\ref{tabnum} lists various values of $\alpha$ with the corresponding limits of the $2\sigma$  range of $\Omega_{c0}$.
\begin{table}[h!]
\centering
\caption{Result of SNe Ia statistical analysis for the time varying vacuum model.}
\centering
\begin{tabular}{ccc}
\hline\hline
$\qquad\alpha\qquad$ & $\qquad\Omega_{c0}\mbox{ (Minimum)}\qquad$ & $\qquad\Omega_{c0}\mbox{ (Maximum)}\qquad$\\ \hline\hline
$\qquad -0.50\qquad$ & $\qquad 0.238 \qquad$ & $\qquad 0.377 \qquad$ \\
$\qquad -0.25\qquad$ & $\qquad 0.207 \qquad$ & $\qquad 0.342 \qquad$ \\
$\qquad -0.05\qquad$ & $\qquad 0.185 \qquad$ & $\qquad 0.316 \qquad$ \\
$\qquad +0.05\qquad$ & $\qquad 0.174 \qquad$ & $\qquad 0.303 \qquad$ \\
$\qquad +0.25\qquad$ & $\qquad 0.154 \qquad$ & $\qquad 0.279 \qquad$ \\
\hline
\end{tabular}
\n{tabnum}
\end{table}
In the following subsection we use the values in TABLE~\ref{tabnum} to
compute the CMB power spectrum.
Afterwards we compare the transfer function for our
time-varying vacuum model obtained with CLASS with the modified BBKS transfer function
proposed in \cite{saulo14,3page}.
Finally, we confront the linear
matter power spectrum with its $\Lambda$CDM counterpart.

\subsection{CMB power spectrum}

In this study we use $h=0.697$ according to TABLE~\ref{tabbg}.
The following steps are similar to those already made in the background analysis.
For each value of $\alpha$ we calculate the CMB
power spectra for the corresponding upper and lower limits of the $2\sigma$ interval for $\Omega_{c0}$
in  TABLE~\ref{tabnum}. Then we compare the resulting plots with the Planck data \cite{planck}.
The results are shown in FIGs.~\ref{cmb1}-\ref{cmb5}. The black solid curves are those for the upper and lower $2\sigma$ limits
of the $\Omega_{c0}$ values, the red solid curve represents  the best-fit Planck result for the $\Lambda$CDM model.  The blue dots are the binned data for the CMB power
spectrum from Planck.
We will consider a time varying vacuum model competitive if the Planck CMB spectrum (blue dots) lies inside  the region which is confined by the two black curves, resulting from the upper and lower limits for $\Omega_{c0}$.
Any model for which the blue dots lie outside this region is ruled out.
FIG.~\ref{cmb1} shows the CMB power spectrum for $\alpha=-0.50$ which can be associated to the model of a cosmological term that decays linearly with the Hubble rate
\cite{Borges,sndec,dec}. Although it reproduces well the position of the first acoustic peak, it clearly does not meet our criterion for a competitive cosmological model.
\begin{figure}[h!]
\includegraphics[width=15cm,height=8cm]{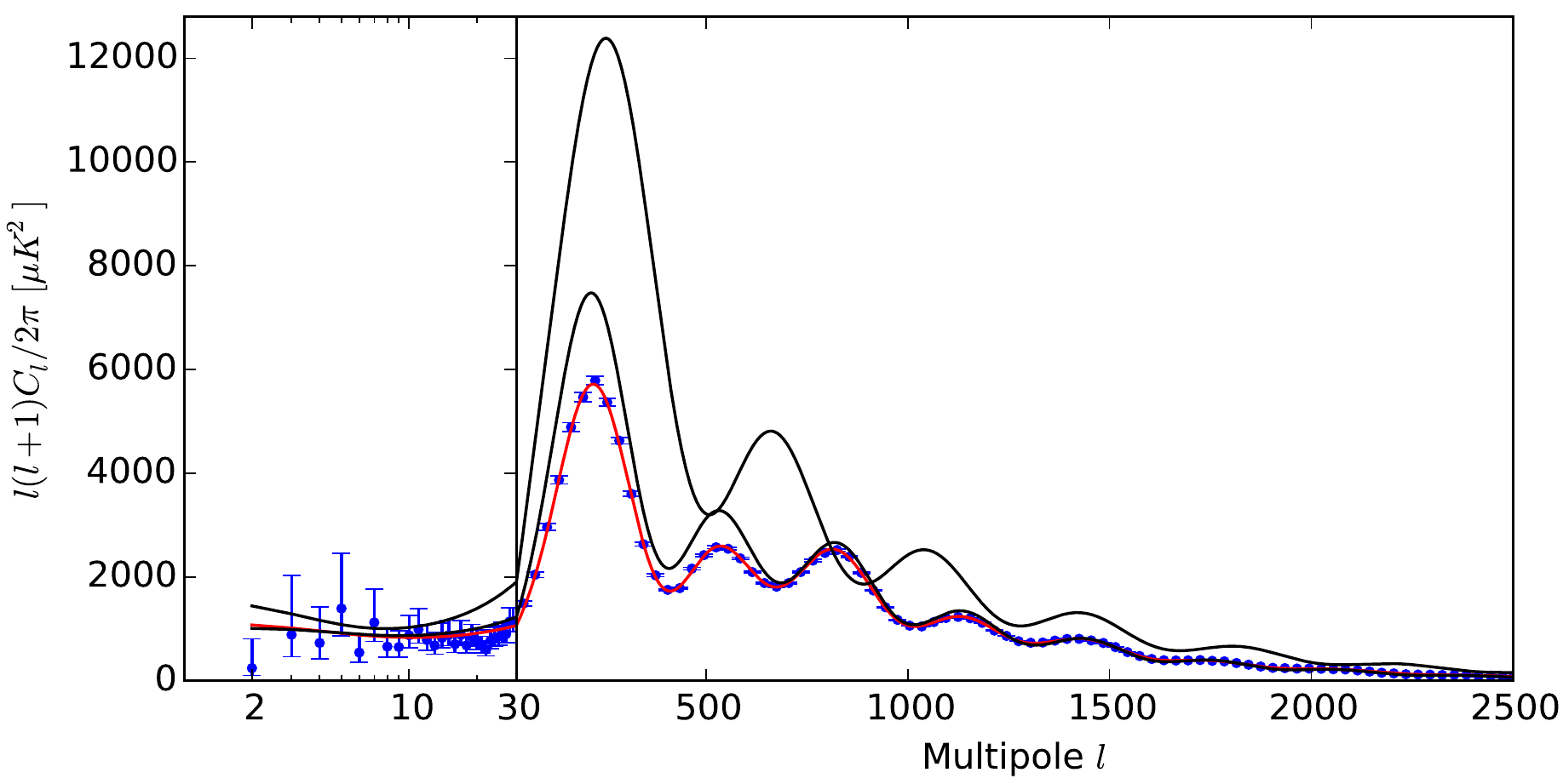}
\caption{CMB spectrum for the decaying vacuum model with $\alpha = -0.50$.}
\n{cmb1}
\end{figure}
There is no concordance between the SNe Ia analysis  and the CMB spectrum. This result confirms an earlier analysis in \cite{xu} and, for small $l$, also reproduces the result of \cite{hermano}.
FIG.~\ref{cmb2} presents the result for $\alpha=-0.25$.
Still, the Planck
result is outside the region between the   black curves although  the difference is somewhat diminished compared with the previous case.
\begin{figure}[h!]
\includegraphics[width=15cm,height=8cm]{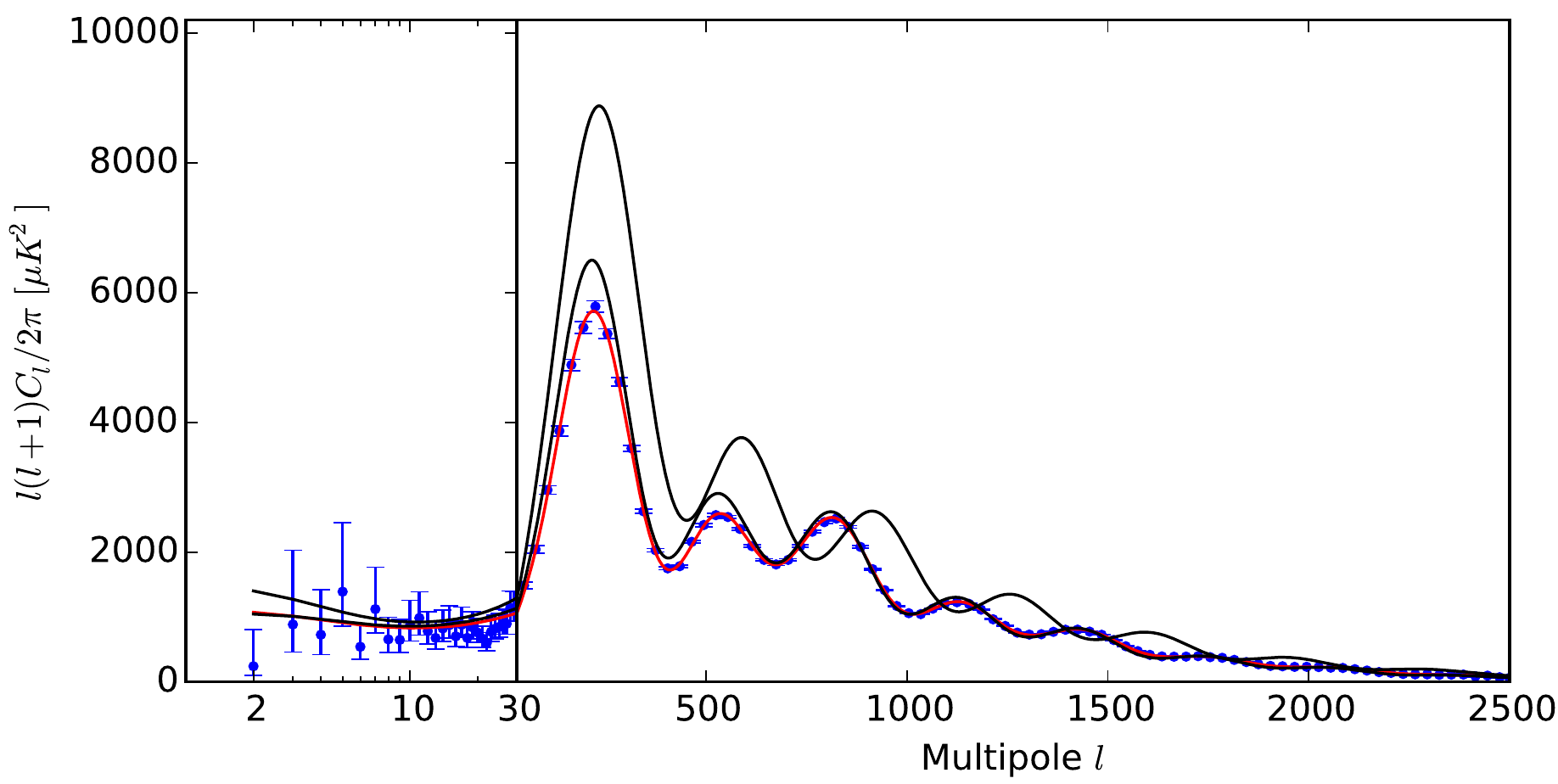}
\caption{CMB power spectrum for the decaying vacuum model with $\alpha = -0.25$.}
\n{cmb2}
\end{figure}
In FIGs.~\ref{cmb3} and \ref{cmb4}, corresponding to $\alpha=-0.05$ and $\alpha=+0.05$
respectively,  the Planck result is inside the regions that are confined by the two black curves which mark the upper and lower limits for $\Omega_{c0}$.
Anticipating that still larger positive values of $\alpha$ are not favored as well,  this means, it is only for $|\alpha|\lesssim 0.05$ that the time varying vacuum models provide a correct CMB spectrum.
In other words, these models have to be very close to the $\Lambda$CDM model.
Recall that a positive $\alpha$ describes an energy transfer from CDM to the vacuum, i.e., a matter decay.
\begin{figure}[h!]
\includegraphics[width=15cm,height=8cm]{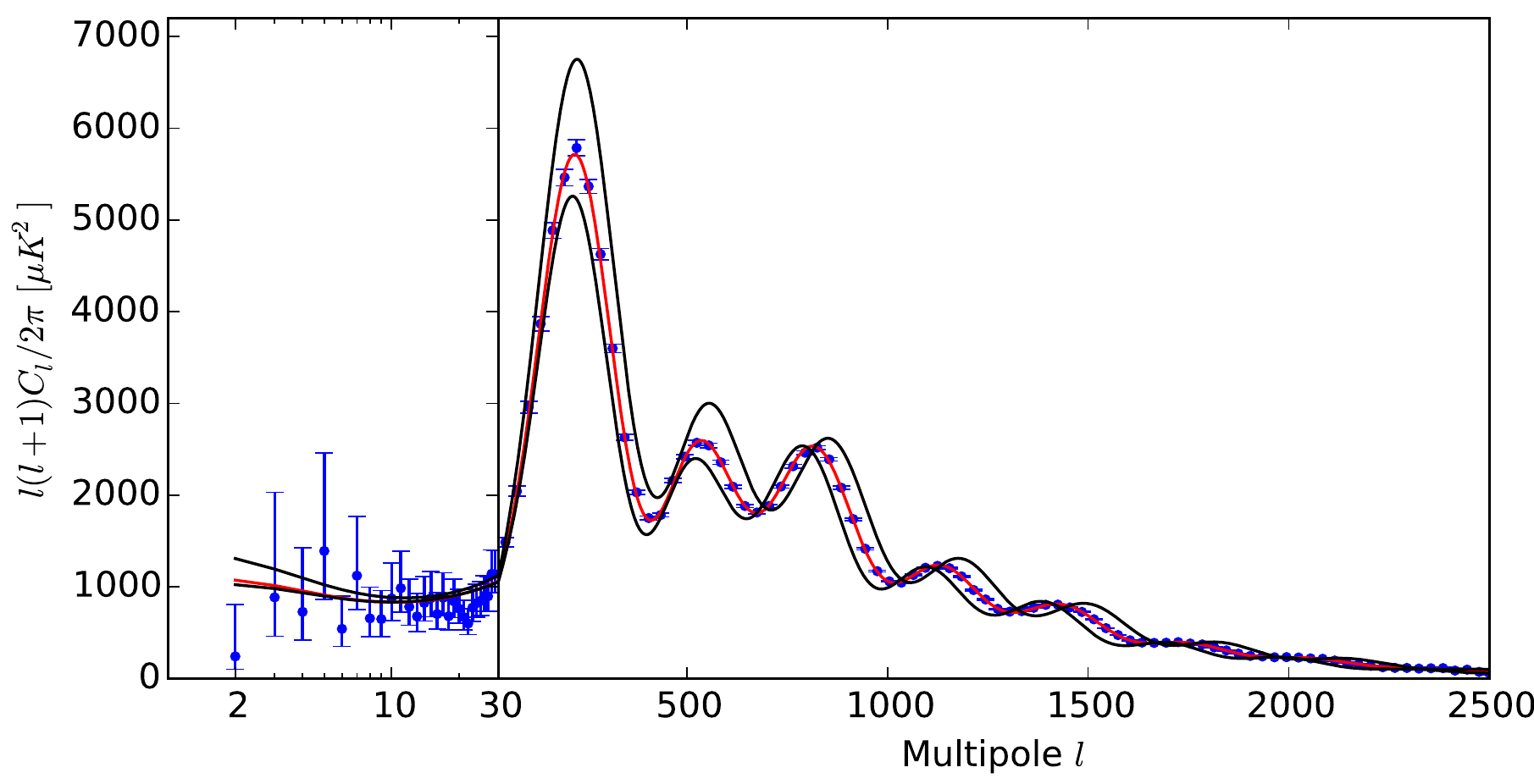}
\caption{CMB power spectrum for the decaying vacuum model with $\alpha = -0.05$.}
\n{cmb3}
\end{figure}
\begin{figure}[h!]
\includegraphics[width=15cm,height=8cm]{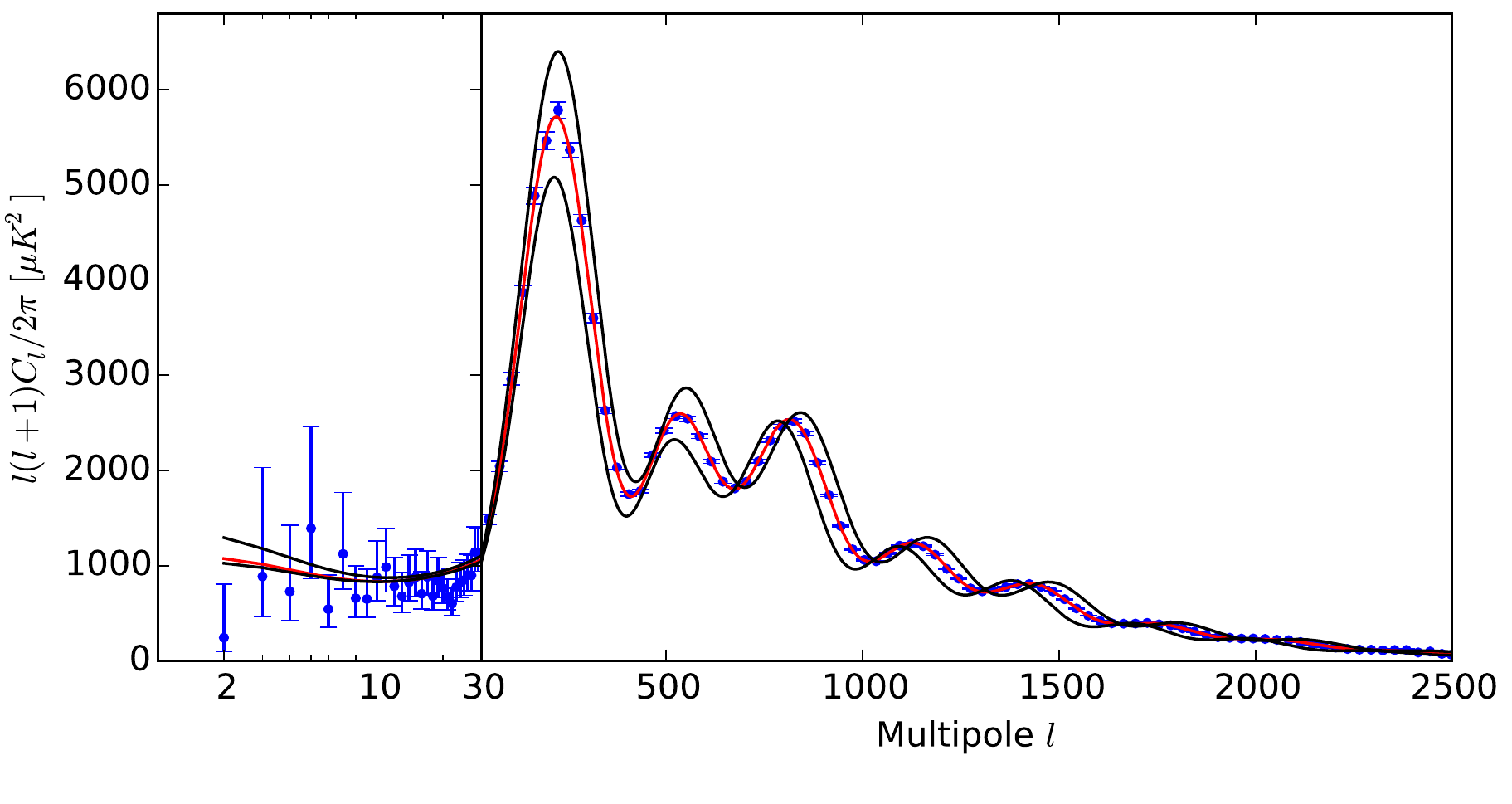}
\caption{CMB power spectrum for the decaying CDM model with $\alpha = +0.05$.}
\n{cmb4}
\end{figure}
The already mentioned result that larger positive values of  $\alpha$ do not correctly describe the
CMB spectrum is visualized in
FIG.~\ref{cmb5}, where we choose $\alpha=+0.25$. In this case the Planck result
is again outside the region confined by the black curves, this time the hight of the first peak is too small, whereas
it was too large in FIGs.~\ref{cmb1} and \ref{cmb2}.
We recall that positive values of $\alpha$ correspond to an energy flux  from the vacuum to CDM.
\begin{figure}[h!]
\includegraphics[width=15cm,height=8cm]{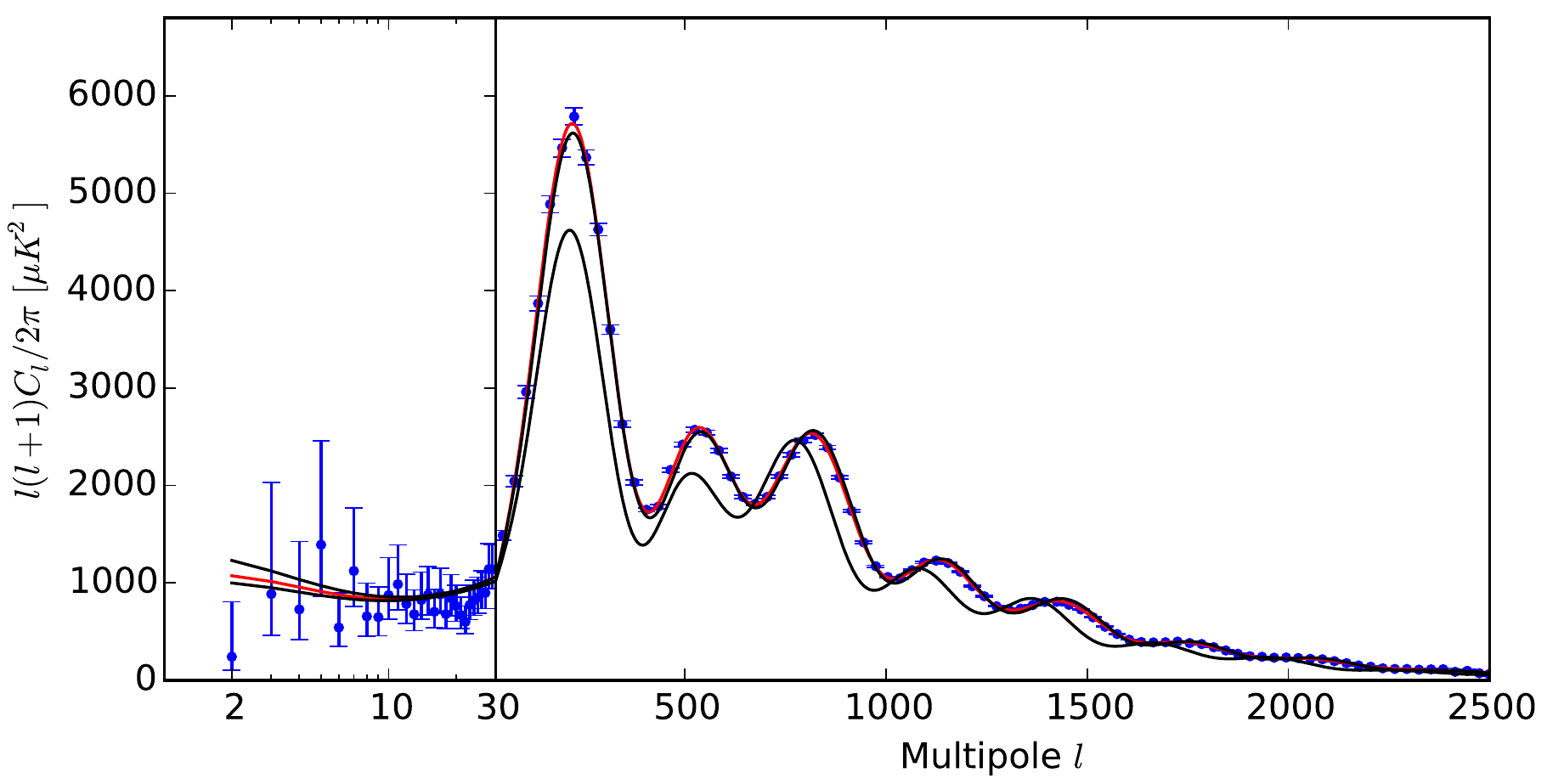}
\caption{CMB power spectrum for the decaying CDM model with $\alpha = +0.25$.}
\n{cmb5}
\end{figure}

\subsection{Transfer function}

Here we consider the matter transfer function for the time-varying vacuum model.
The transfer function is generally defined as (cf. \cite{EH97})
\begin{equation}\label{T}
T(k)\equiv \frac{\delta_{m}\left(k, z=0\right)}{\delta_{m}\left(k, z=\infty\right)}\frac{\delta_{m}\left(0, z=0\right)}{\delta_{m}\left(0, z=\infty\right)}.
\end{equation}
The widely used BBKS transfer function \cite{bbks} is known to be a
fit formula for the $\Lambda$CDM
model that depends on $k/k_{eq}$, where $k_{eq}$ is the comoving
horizon scale at the time of matter-radiation equality,
\be
T\left(x=k/k_{eq}\right)=\dfrac{\ln\left(1+0.171 x\right)}{\left(0.171 x\right)}\left[1+0.284 x+\left(1.18 x\right)^{2}+\left(0.399\right)^{3}+\left(0.490 x\right)^{4}\right]^{-0.25}.
\n{BBKS}
\ee
We start by comparing the transfer functions from CLASS \cite{class}, from Eisenstein and Hu \cite{EH97} (EH97) and from \cite{bbks} (BBKS86) for the $\Lambda$CDM model. This is shown in FIG.~\ref{transferLambda}.
\begin{figure}[h!]
\includegraphics[scale=0.63]{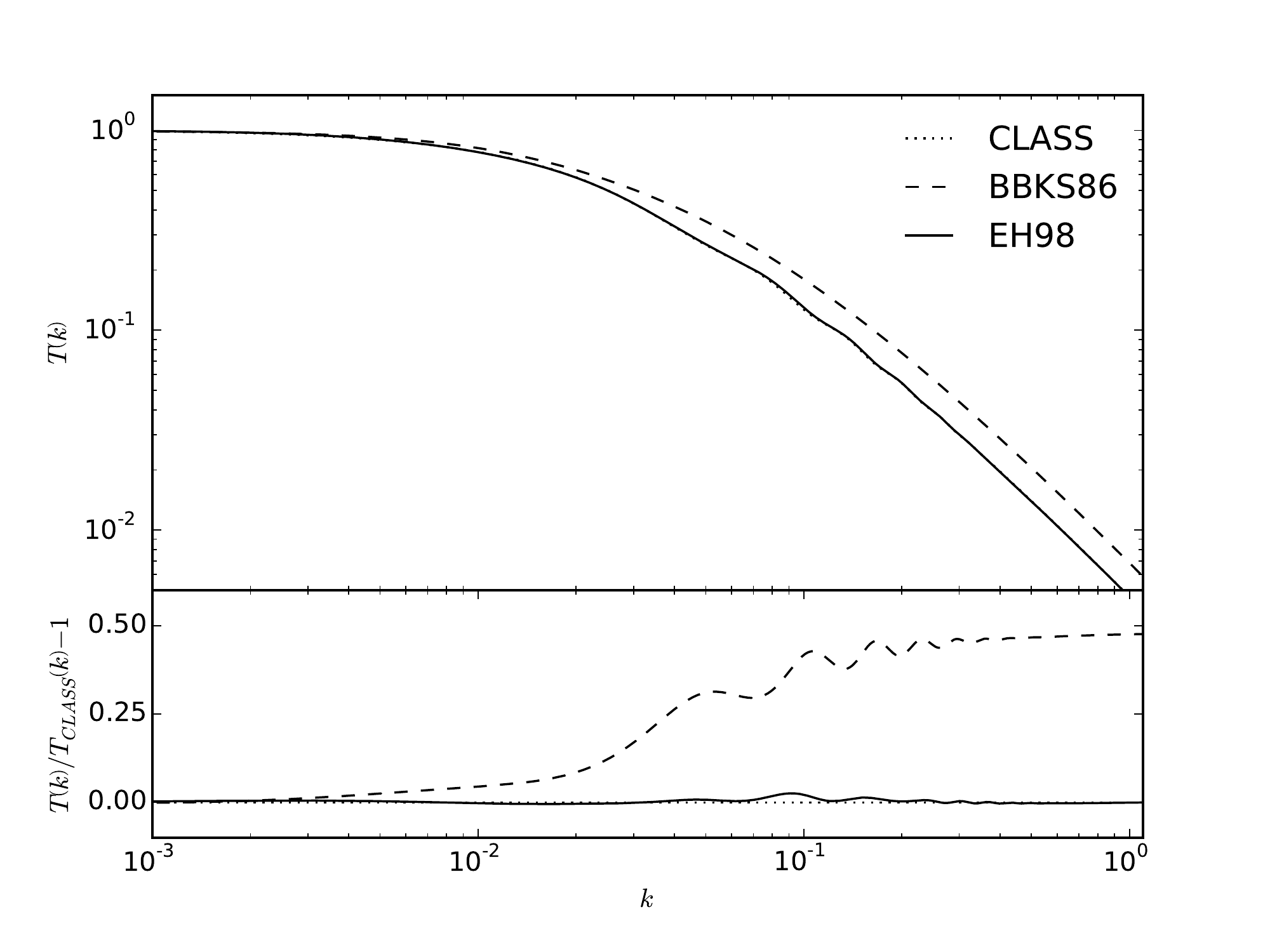}
\caption{Transfer functions for the $\Lambda$CDM model.
BBKS86 refers to \cite{bbks} and EH97 refers to \cite{EH97}.
The curves EH97 and CLASS are almost indistinguishable in the upper part.
The lower part visualizes the departure of the BBKS transfer function from the transfer function from CLASS for $k\gtrsim 10^{-2}$, i.e. still in the linear regime while the curves for CLASS and EH97 almost coincide in the entire range.}
\n{transferLambda}
\end{figure}
A different representation of the same curves with a higher resolution and with the inclusion of baryons in the EH97 transfer functions is provided in
FIG.~\ref{transLambda2}.  Here, $f_{b}= \frac{\Omega_{b0}}{\Omega_{m0}}$.
The differences between the BBKS86 and the EH97 curves are of the order of 5\% only if baryons are not included.
Otherwise there are substantial deviations.
On the other hand, the EH97 transfer function with baryons ($f_{b}= 0.158$) reproduces the result from CLASS to better than 3\% (lower part of the figure).
\begin{figure}[h!]
\includegraphics[scale=0.63]{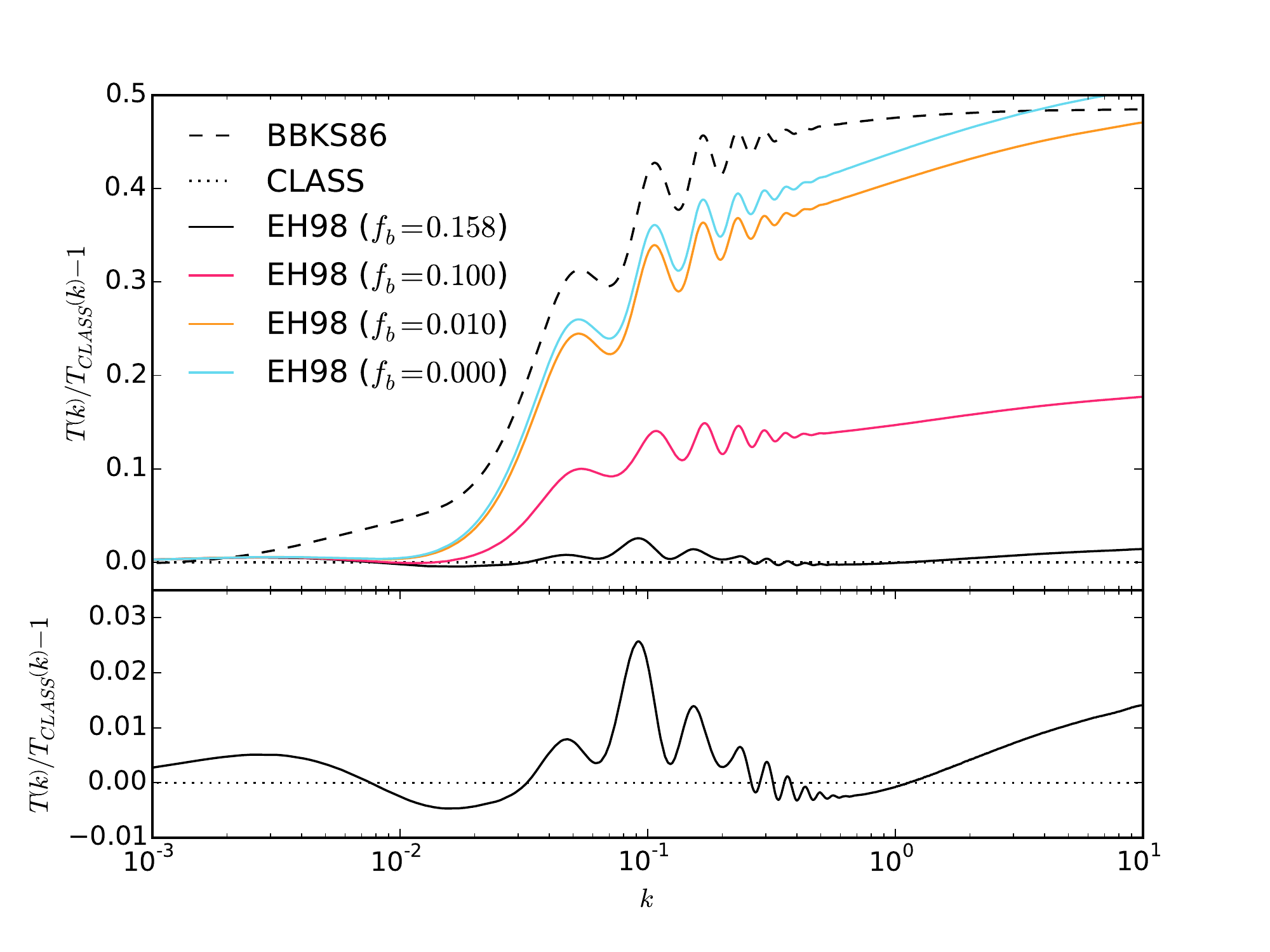}
\caption{Transfer functions for the $\Lambda$CDM model. The upper part depicts the deviations of the CLASS and of the EH97 transfer functions with different baryon content from BBKS86. The latter is a reasonable approximation (at the 5\% level) to EH97 only if baryons are not taken into account. EH97 with $f_{b}= 0.158$ coincides with CLASS up to 3\% (lower part of the figure).}
\n{transLambda2}
\end{figure}
It has been suggested that the linear matter power
spectrum of a decaying vacuum model can be obtained using a modified BBKS
transfer function \cite{saulo14,3page}.
The modification is motivated by the fact that matter production will change
the time of matter-radiation equality compared with the $\Lambda$CDM model.
According to \cite{saulo14,3page} this is taken into account by a modification in $k_{eq}$ in (\ref{BBKS}),
\be
k_{eq}=\sqrt{\dfrac{2}{\Omega_{r0}}}\dfrac{\Omega_{c0}^{\frac{1}{1+\alpha}}}{l_{H0}},
\n{deckeq}
\ee
where $l_{H0}$ is the present Hubble radius. The $\Lambda$CDM value corresponds to $\alpha = 0$.
In FIG.~\ref{transdec} we compare this modified BBKS transfer
function with the CLASS transfer function for the best-fit values in TABLE~\ref{tabbg}.
For comparison we have also included the EH97 transfer function with the modification (\ref{deckeq}).
The modification (\ref{deckeq}) is indicated by a plus sign.
It is obvious that the modified BBKS function is considerably different from its CLASS counterpart for $k\gtrsim 10^{-2}$.
\begin{figure}[h!]
\includegraphics[scale=0.63]{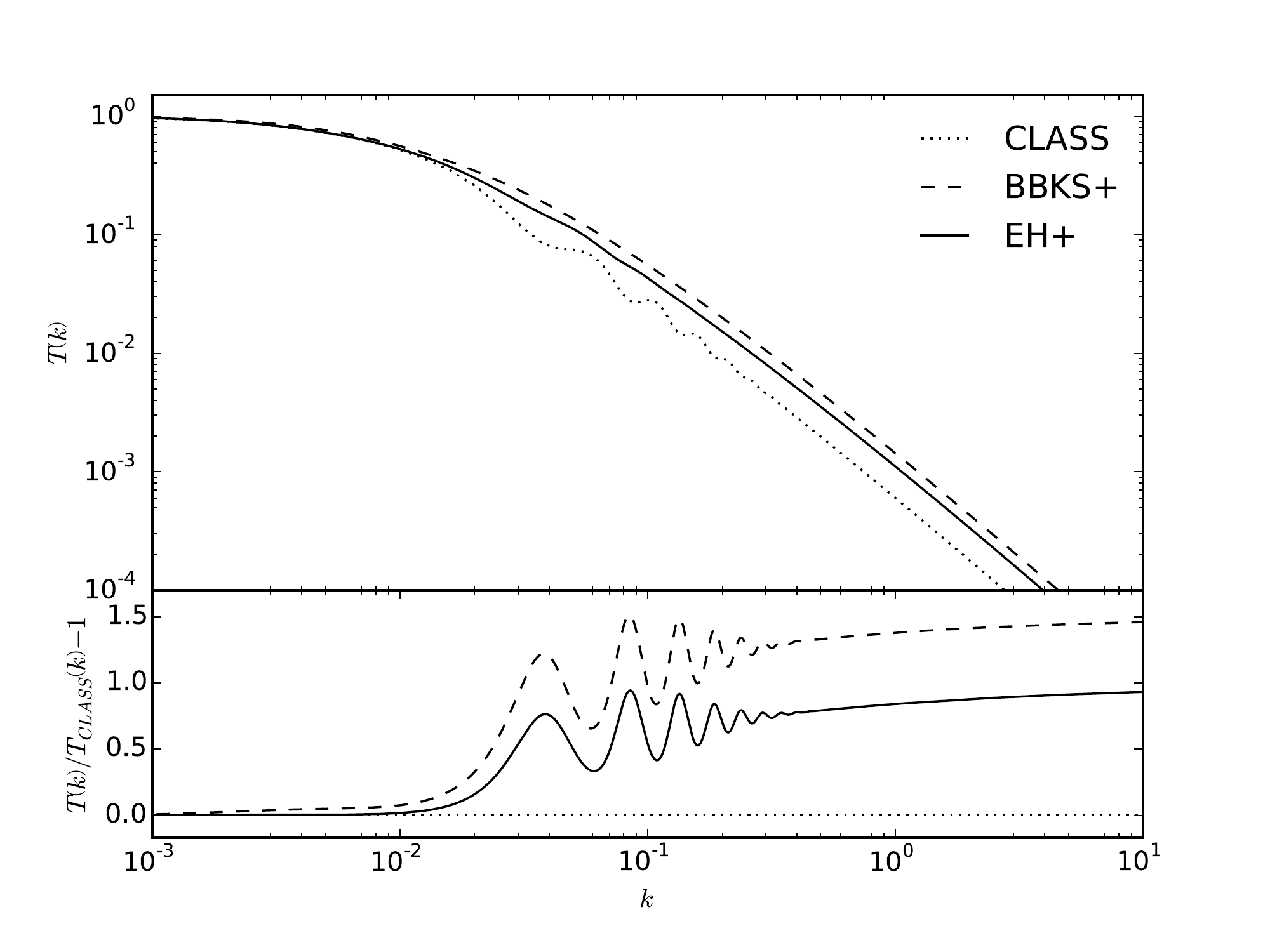}
\caption{Transfer functions for the decaying vacuum model with the best-fit values of TABLE~\ref{tabbg}.
The plus sign in BBKS+ and EH+ indicates that in the transfer functions \cite{bbks} (see (\ref{BBKS})) and \cite{EH97}
the modification (\ref{deckeq}) was applied. For any $k\gtrsim 10^{-2}$ this does not result in an acceptable approximation to the result from CLASS.}
\n{transdec}
\end{figure}

\subsection{Growth rate and linear matter power spectrum}

The modified CLASS code also allows us to obtain the growth rate and the linear
matter power spectrum for the time-varying vacuum model.
The growth rate is defined by $f = \frac{d\ln \delta_{m}}{d\ln a}$.
Both $\delta_{m}$ and $f$ are depicted in FIG.~\ref{figrowth} for the values $\alpha = -0.5$, $\alpha = -0.25$, $\alpha = -0.05$, $\alpha = +0.05$ and $\alpha = +0.25$, listed in TABLE~\ref{tabnum}.
In FIG.~\ref{figmatter} we illustrate
 the corresponding power spectra for all these cases.
The black solid
lines are obtained analogously to those of the previous CMB analysis.
A chosen $\alpha$ is combined with the upper and lower limits of the admissible range of
$\Omega_{c0}$ values.
The red lines represent the $\Lambda$CDM results.
Consistent with the results for the CMB spectra, it is only for $\alpha=-0.05$ and $\alpha=+0.05$ that
the $\Lambda$CDM curves in FIGs.~\ref{figrowth} and \ref{figmatter}  lie inside the range limited by the combination of $\alpha$ with the admitted
values for $\Omega_{c0}$. This double confirmation gives strong support to the conclusion that these types of time-varying vacuum models are strongly constrained. They survive only as long as they stay very close to the standard model.

\begin{figure}[h!]
\subfigure{
\includegraphics[width=6.5cm,height=6.5cm,keepaspectratio]{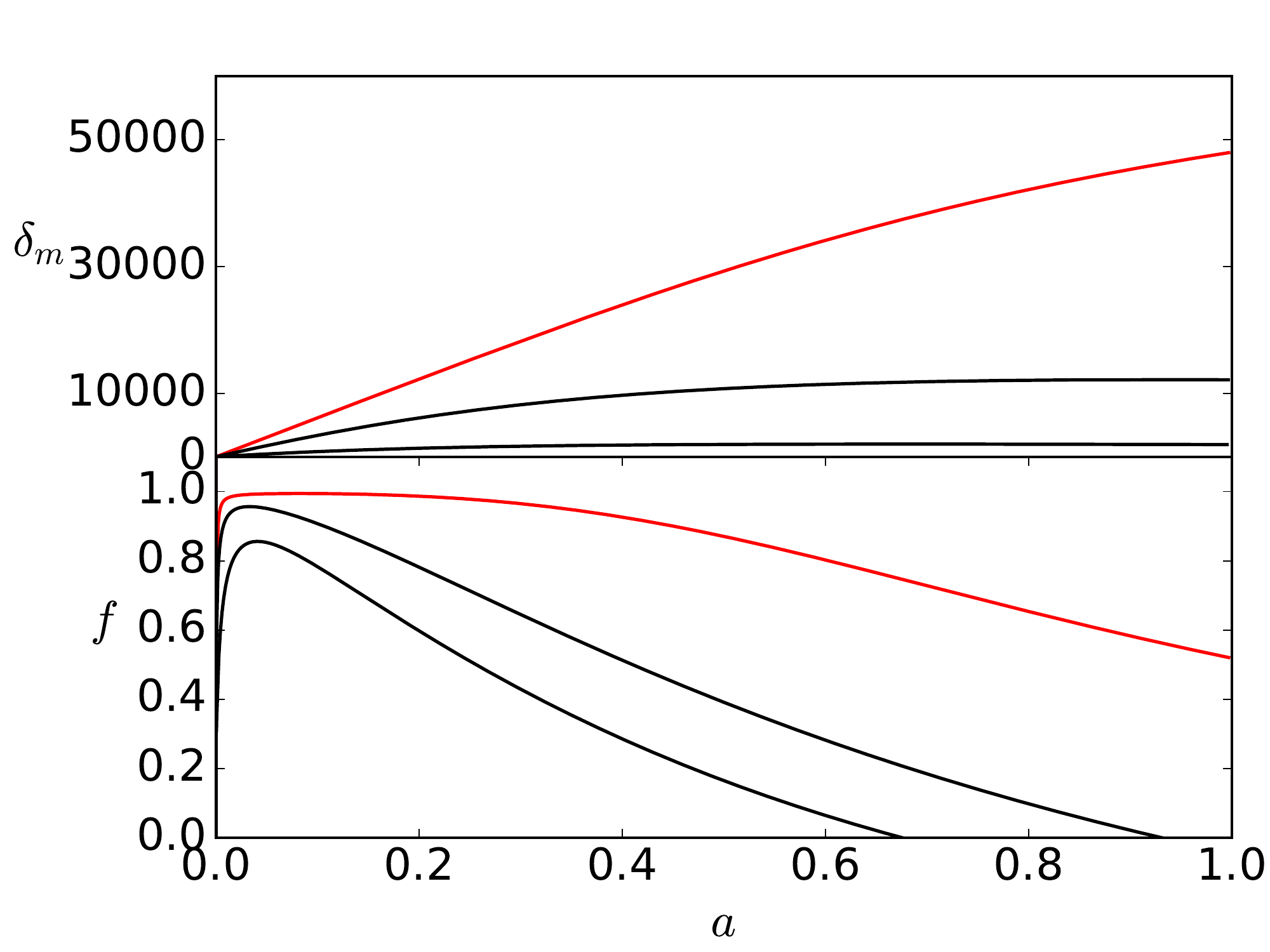}}
\subfigure{
\includegraphics[width=6.5cm,height=6.5cm,keepaspectratio]{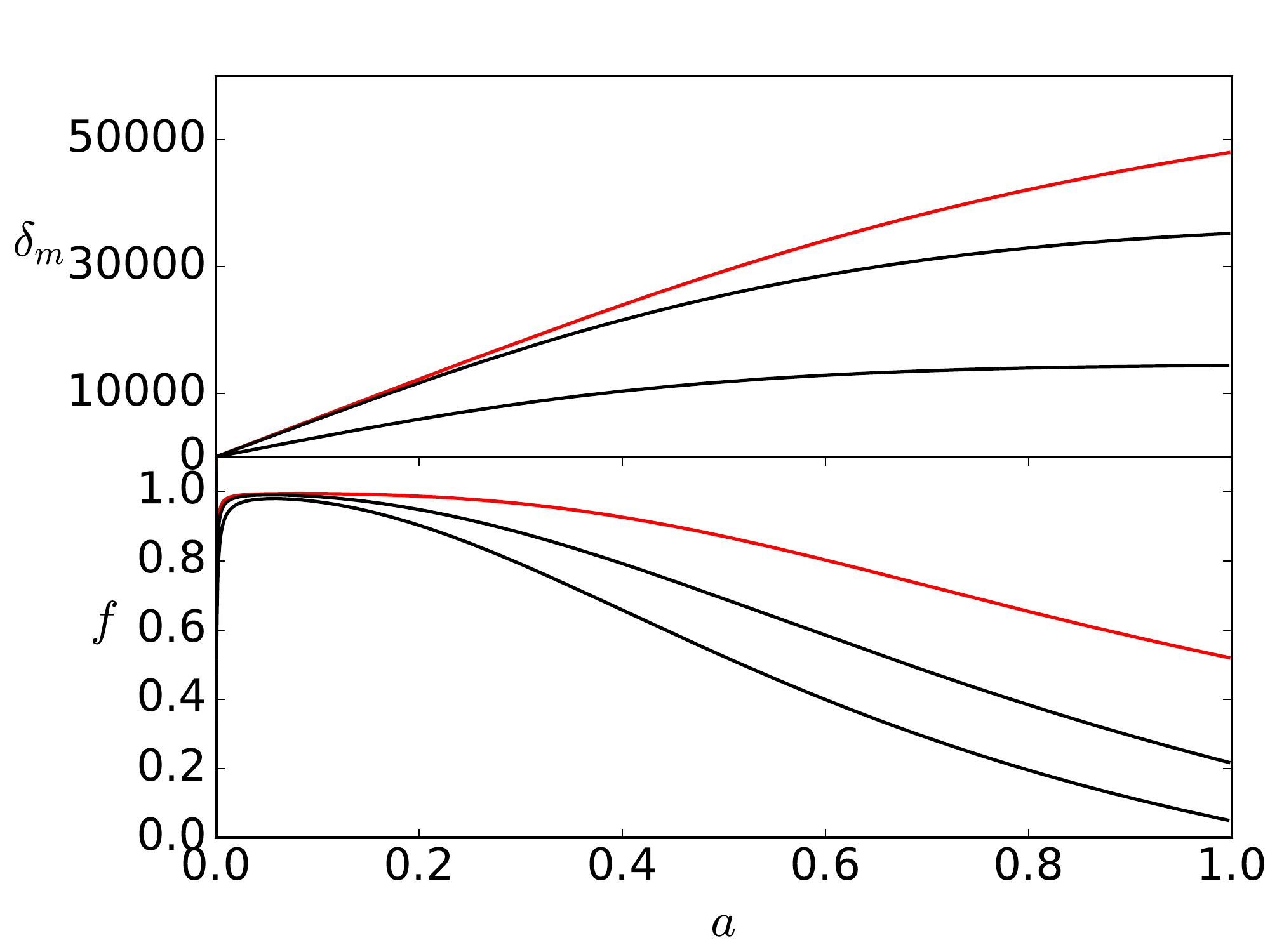}}\\
\subfigure{
\includegraphics[width=6.5cm,height=6.5cm,keepaspectratio]{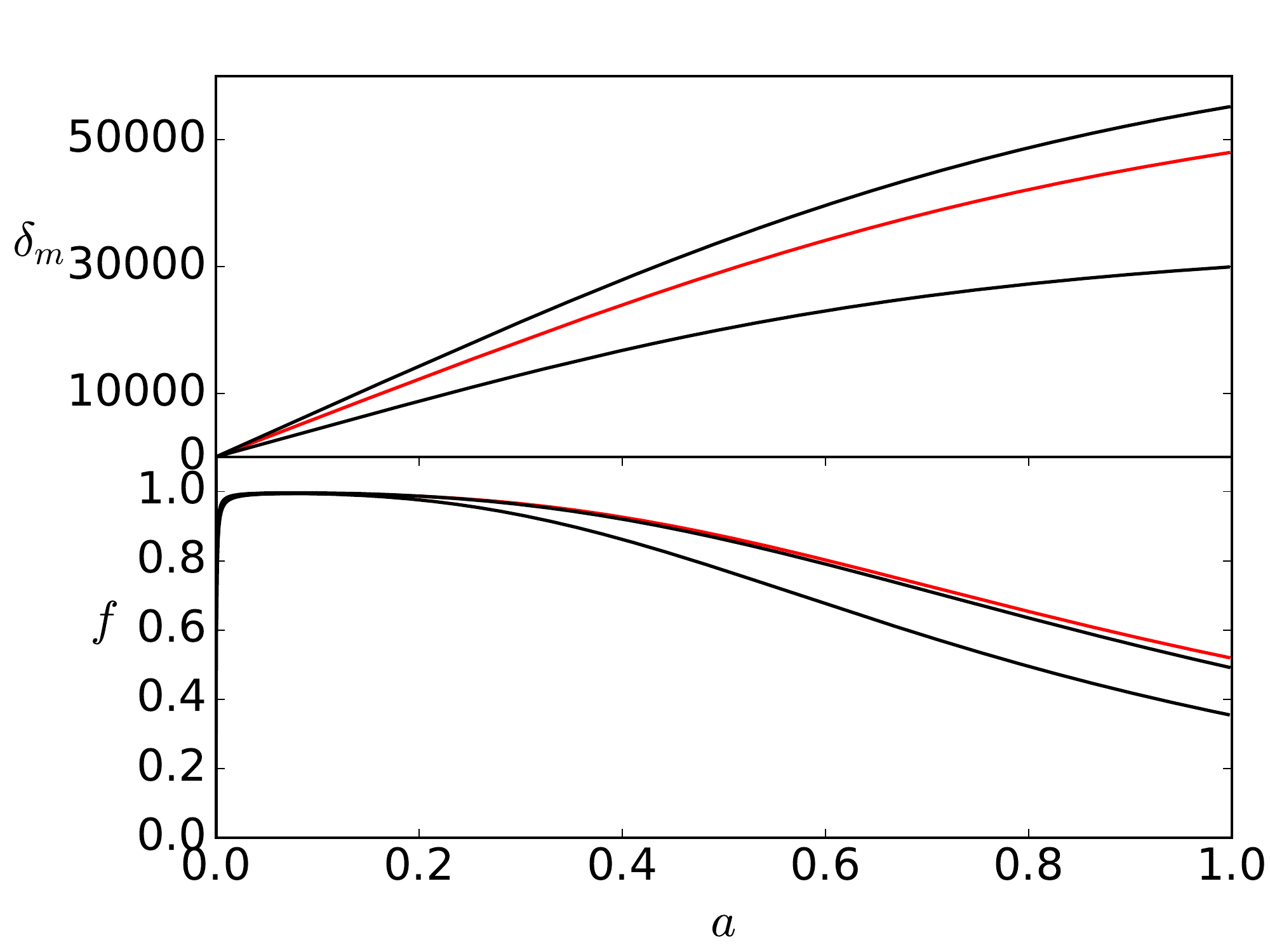}}
\subfigure{
\includegraphics[width=6.5cm,height=6.5cm,keepaspectratio]{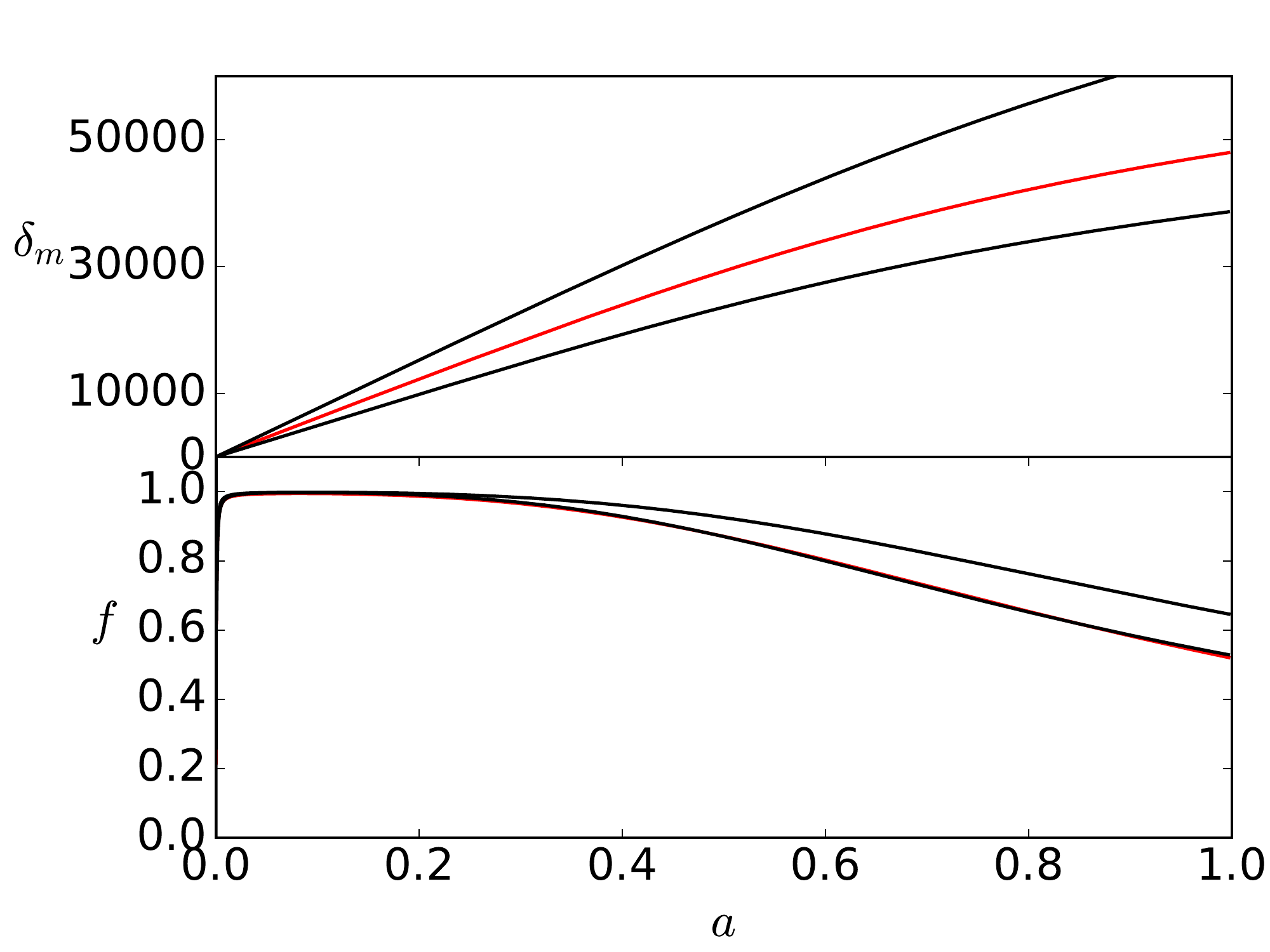}}\\
\subfigure{
\includegraphics[width=6.5cm,height=6.5cm,keepaspectratio]{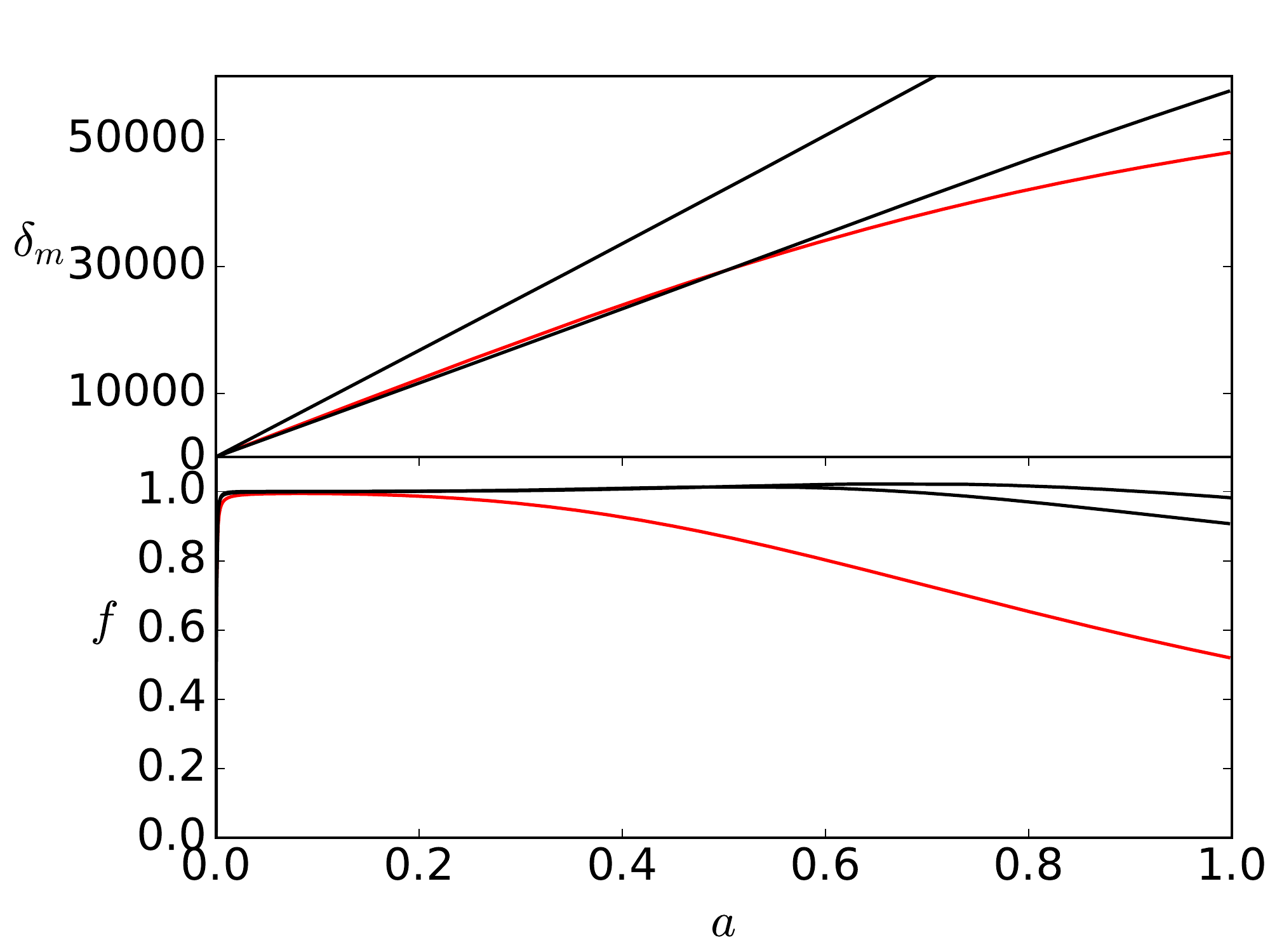}}
\caption{Matter density contrast $\delta_{m}$  and growth rate $f$ for the parameters of TABLE~\ref{tabnum}.
Upper part: $\alpha = -0.5$, $\alpha = -0.25$ (from left to right). Middle: $\alpha = -0.05$, $\alpha = +0.05$. Lower part:  $\alpha = +0.25$.}
\n{figrowth}
\end{figure}

\begin{figure}[h!]
\includegraphics[scale=0.63]{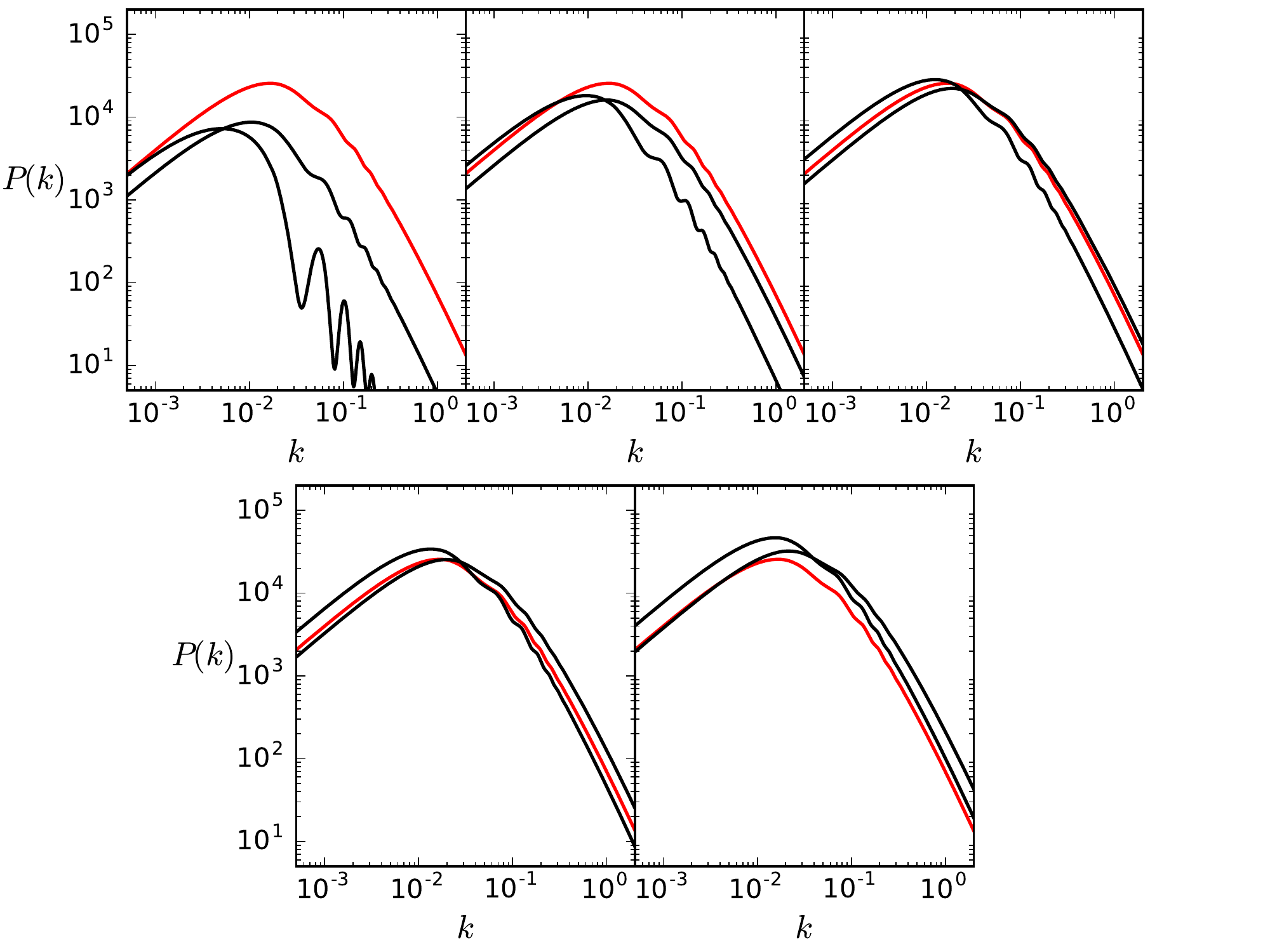}
\caption{Linear matter power spectrum for the parameters of TABLE~\ref{tabnum}. Upper part: $\alpha = -0.5$, $\alpha = -0.25$, $\alpha = -0.05$ (from left to right). Lower part: $\alpha = +0.05$ and $\alpha = +0.25$.}
\n{figmatter}
\end{figure}

\section{Conclusions}
\label{summary}

A cosmological dynamics in which the dark sector is modeled as a gCg with an EoS $p=-\frac{A}{^{\rho\alpha}}$ is only compatible with observations if the parameter $\alpha$ is restricted to $|\alpha|\lesssim 0.05$, i.e., it has to be very close to the $\Lambda$CDM model which corresponds to $\alpha =0$. A negative $\alpha$ describes a decaying vacuum, whereas $\alpha >0$ is equivalent to a decay of dark matter.
While the SNIa analysis on the basis of the JLA sample leaves room for a broad range of date including $\alpha =-1 $ and $\alpha =0$ (at the $2\sigma$ confidence level), the Planck data for the CMB anisotropy spectrum narrow the admissible interval drastically.
The limits obtained from a comparison of the matter power spectrum of the gCg-based model with the corresponding spectrum of the standard model are consistent with $|\alpha|\lesssim 0.05$ as well.
We demonstrate that the mere position of the first acoustic peak in the CMB spectrum is not sufficient to assess a cosmological model. In particular, our study is incompatible with a model in which vacuum energy decays linearly with the Hubble rate, corresponding to $\alpha =-\frac{1}{2}$.
Further, we point out that the BBKS matter transfer function does not provide a good approximation to the transfer function of the CLASS code if baryons are taken into account. Modifications of the BBKS expression do not lead to acceptable results.
While there remains a small range of admissible $\alpha$ values around zero, our analysis may well be seen as a confirmation of the standard $\Lambda$CDM model.

\ \\
\textbf{Acknowledgment}\\
Support by CNPq, CAPES and FAPES is gratefully acknowledged. DFM acknowledges the support of the Norwegian Research Council. LC thanks O. Piattella for useful discussions.

\end{document}